\documentclass{ametsocV6.1}

\usepackage{multirow}
\usepackage{caption}
\nolinenumbers

\title{Causality for Earth science - A Review on Time-series and Spatiotemporal Causality Methods}

\authors{Sahara Ali,\aff{a}\aff{b}\aff{c} 
Uzma Hasan,\aff{a} 
Xingyan Li,\aff{a} 
Omar Faruque,\aff{a} 
Akila Sampath, \aff{a}\aff{b}
Yiyi Huang, \aff{a}\aff{b}
Md Osman Gani,\aff{a}\aff{b} 
and Jianwu Wang,\aff{a}\aff{b}\correspondingauthor{Jianwu Wang, jianwu@umbc.edu}
}

\affiliation{\aff{a}{Department of Information Systems, University of Maryland Baltimore County, Baltimore, MD 21228, USA}\\
\aff{b}{NSF HDR Institute for Harnessing Data and Model Revolution in the Polar Regions (iHARP), United States}\\
\aff{c}{Department of Information Science, University of North Texas, Denton, TX 76203, USA}\\
}

\abstract{
This survey paper covers the breadth and depth of time-series and spatiotemporal causality methods,  and their applications in Earth science. The paper first introduces the concepts of causal discovery and causal inference, followed by the underlying causal assumptions, evaluation techniques and key terminologies of the two domain areas. The paper elicits the various state-of-the-art methods introduced for time-series and spatiotemporal causal analysis along with their strengths and limitations. The paper further describes the existing applications of several methods for answering specific Earth science questions such as extreme weather events, sea level rise, teleconnections, etc. Our survey paper will benefit the Earth science community interested in taking an AI-driven approach to study the causality of different dynamic and thermodynamic processes as we present the open challenges and opportunities in performing causality-based Earth science study. It will also serve as a primer for data science researchers interested in data-driven causal study as we share a holistic list of resources, such as Earth science datasets (synthetic, simulated and observational data) and open source tools for causal analysis. }

\begin{document}
\maketitle

\section*{Significance Statement}
Causality is the study of discovering cause-effect relations in data. Earth science can benefit greatly from causal methods as researchers have started utilizing causality to understand the complex interactions leading to climate change, long term weather patterns, extreme weather events, etc. This paper provides a primer to causal inference and causal discovery for different Earth science domains with some of the existing applications and open questions for future research. The paper also comprises a holistic list of available tools and datasets that can help both data science and Earth science communities in performing causal analysis on Earth data.
\section{Introduction}
Earth science is a broad field that studies the Earth's composition, processes, and physical characteristics \citep{smith2013does}. It encompasses several sub-disciplines like geology (rocks and minerals), meteorology (atmosphere and weather), oceanography (oceans and their properties), and environmental science (interactions between living things and the environment).  Climate models, complex computer simulations, play a vital role in Earth science research. These models enable predictions about climate trends and allow for the testing of various scenarios, such as analyzing the global carbon footprint in order to mitigate greenhouse gas emissions. Despite their importance, climate models are computationally demanding and necessitate significant computing resources. This requirement stems from the need to simulate numerous dynamic processes occurring in the atmosphere, oceans, and on land, such as cloud formation and ocean current generation.

 In recent decades, there has been a significant increase in the availability of large-scale climate data from various observational sources (such as satellite remote sensing, ground-based, sea-borne, or air-borne systems) and climate model outputs \citep{guo2015earth}. This large volume of data has opened up new ways to use data-driven methods for observational causal discoveries without relying on the correlation and trend analyses of the data \citep{rubin2005causal}. Causal structure discovery (CSD) models are becoming increasingly valuable across various research areas in Earth system sciences  \citep{melkas2021interactive}. Climate causal studies pose the changing trend of applying the current state-of-the-art climate data not only to correlation and regression methods but also to causal inference methods. For instance, climate researchers have realized that climate simulations introduce ambiguous values to the datasets which makes them not applicable in decision-making applications \citep{CausalDiscoveryforClimateResearchUsingGraphicalModels}. After understanding the importance of data-driven methods in climate science, scientists have started to explore different causal structure algorithms and Bayesian networks to get deeper insights into causal hypotheses and the evaluation of physical models \citep{CausalDiscoveryforClimateResearchUsingGraphicalModels}. 
 To sum it up, causality methods can play a crucial role in Earth science by helping researchers understand and quantify the cause-and-effect relationships within complex Earth systems for a variety of usecases \citep{nowack_causal_2020}. Some of these include: identifying the primary drivers of environmental changes, attributing specific climate events like volcanic eruptions, identifying causes of extreme weather events such as hurricanes, droughts and floods, helping in the prediction of natural hazards such as earthquakes and tsunamis, designing effective pollution control policies, understanding El Niño/La Niña events and their global climate impacts, etc. 

This survey paper presents a comprehensive overview of the current understanding of causality in Earth science. It serves as an introduction to key concepts in causality and the commonly used methods in this field. Additionally, this survey identifies gaps in current research and highlights areas where causal methods could improve our understanding of the Earth's system. The paper is organized as follows: Section 2 discusses the open challenges in causality-based studies of Earth science problems. Section 3 explores the concept of causal structure learning, or "Causal Discovery". This section covers foundational concepts such as key terms, causal assumptions, and relevant evaluation metrics for structure learning. It also details various approaches for performing causal discovery on time-series and spatiotemporal data and reviews their applications in Earth science. Section 4 delves into "Causal Inference" techniques for estimating causal effects. It also covers state-of-the-art methods for time-series and spatiotemporal causal inference, detailing their strengths and limitations and providing an example of their application in Earth science. Section 5 lists resources for synthetic, simulated, and observational data commonly used in Earth science, as well as open-source toolboxes for causal analysis. Section 6 wraps up the paper with a discussion on potential research directions.
We aim for this paper to benefit both the Data Science community, which is interested in data-driven causal studies, and the Earth science community, which seeks to use AI-driven approaches to explore the causality of various dynamic and thermodynamic processes.

\begin{landscape}
\begin{table}[ht!]
\centering
\caption{Comparison among the existing survey papers.}
\label{tab:my-table}
\begin{tabular}{llllllll}
\toprule
\multicolumn{1}{|c|}{Paper Title} & \multicolumn{1}{c|}{Discovery} & \multicolumn{1}{c|}{Inference} & \multicolumn{1}{c|}{Datasets} & \multicolumn{1}{c|}{Metrics} & \multicolumn{1}{c|}{Software} & \multicolumn{1}{c|}{Time series} & \multicolumn{1}{c|}{Spatial} \\ \midrule
Inferring causation from time series in Earth system sciences \citep{runge_inferring_2019} & $\checkmark$ &  \xmark & $\checkmark$ & \xmark & \xmark & $\checkmark$ &  \xmark\\ 
D’ya Like DAGs? A Survey on Structure Learning and Causal Discovery \citep{vowels2022d} & $\checkmark$ & \xmark & $\checkmark$ & $\checkmark$ & $\checkmark$ & $\checkmark$ & \xmark \\ 
Survey and Evaluation of Causal Discovery Methods for Time Series \citep{assaad2022survey} & $\checkmark$ & \xmark & $\checkmark$ & $\checkmark$ & \xmark & $\checkmark$ & \xmark \\ 
Causal Structure Learning \citep{heinze2018causal} & $\checkmark$ & \xmark & $\checkmark$ & $\checkmark$ & \xmark & \xmark & \xmark \\ 
Review of Causal Discovery Methods Based on Graphical Models \citep{glymour2019review} & $\checkmark$ & \xmark & \xmark & \xmark & \xmark & $\checkmark$ & \xmark \\ 
A Survey of Learning Causality with Data: Problems and Methods \citep{guo2020survey} & $\checkmark$ & $\checkmark$ & $\checkmark$ & $\checkmark$ & $\checkmark$ & $\checkmark$ & \xmark \\ 
Causal inference for time series analysis: problems,methods and evaluation \citep{moraffah2021causal} & $\checkmark$ & $\checkmark$ & $\checkmark$ & $\checkmark$ & \xmark & $\checkmark$ & \xmark \\ 
A Survey on Causal Discovery Methods for I.I.D. and Time Series Data \citep{hasan2023survey} & $\checkmark$ & \xmark & $\checkmark$ & $\checkmark$ & $\checkmark$ & $\checkmark$ & \xmark \\ 
Causal Inference for Time Series \citep{runge2023causal} & $\checkmark$ & $\checkmark$ & \xmark & $\checkmark$ & $\checkmark$ & $\checkmark$ &  \xmark \\
A Primer on Deep Learning for Causal Inference \citep{koch2021deep} & \xmark & $\checkmark$ & \xmark & $\checkmark$ & $\checkmark$ & $\checkmark$ &  \xmark \\
Causal Discovery from Temporal Data: An Overview and New Perspectives \citep{gong2023causal} & $\checkmark$ & \xmark & $\checkmark$ & $\checkmark$ & \xmark & $\checkmark$ & \xmark \\ 
A Survey on Causal Inference \citep{yao2021survey} & \xmark & $\checkmark$ & \xmark & $\checkmark$ & $\checkmark$ & \xmark & \xmark \\ 
Causal inference for process understanding in Earth science s \citep{massmann2021causal} & $\checkmark$ & $\checkmark$ & \xmark & \xmark & \xmark & \xmark & \xmark \\ 
Spatial Causality: A Systematic Review on Spatial Causal Inference \citep{akbari2023spatial} & \xmark & $\checkmark$ & \xmark & $\checkmark$ & \xmark & \xmark & $\checkmark$ \\
A review of spatial causal inference methods for environmental and epidemiological applications \citep{reich2021review} & $\checkmark$ & $\checkmark$ & \xmark & $\checkmark$ & \xmark & \xmark & $\checkmark$ \\ 
Our Survey & $\checkmark$ & $\checkmark$ & $\checkmark$ & $\checkmark$ & $\checkmark$ & $\checkmark$ & $\checkmark$ \\ \bottomrule
\end{tabular}
\end{table}
\end{landscape}

\begin{table}[ht!]
\centering
\caption{Examples of causality-related open questions in Earth science domains.}
\label{tab:earth_chlng}
\begin{tabular}{ll}
\toprule
\textbf{Earth science Domain} & \textbf{Key Questions}\\ 
\midrule
Atmosphere & How do natural and anthropogenic activities influence greenhouse gas concentrations and atmospheric composition?\\
 & How do aerosols influence cloud formation, precipitation patterns, and regional climate variability?\\
 & How are climate change and global warming influencing patterns of weather extremes, including heatwaves, \\
 & droughts, extreme rainfall events, and storms?\\
 & What are the mechanisms driving changes in atmospheric circulation patterns, such as shifts in jet streams \\
 & and tropical cyclones?\\
 Cryosphere & What are the primary drivers of amplified Arctic warming and accelerated loss of sea ice?\\
 & How does Arctic amplification contribute to more frequent extreme heat, wildfire and increasing precipitation \\
 & at high latitudes?\\
 & What factors drive the loss of glacier mass in both the Arctic and Antarctic regions, and how does it\\
 & contribute to global weather patterns?\\
 Hydrosphere & What are the climatic and non-climatic drives for water cycle change?\\
 & How does sea level rise impact extreme events such as, floods and droughts?\\
 & What are the effects of alterations in soil moisture on local weather patterns, water security, and agricultural systems?\\
 Ocean & How are rising global temperatures affecting ocean temperatures, both at the surface and in deeper layers?\\
 & How is increased atmospheric carbon dioxide leading to ocean acidification?\\
 & How is climate change affecting ocean circulation patterns, including the Gulf Stream and the Atlantic \\
 & Meridional Overturning Circulation (AMOC)?\\
 & What are the feedback loops between changes in the ocean carbon cycle and global climate dynamics?\\
 Biosphere & How is rising global temperature influencing the geographic ranges and habitats of species?\\
 & How does amplified Arctic warming affect polar biodiversity?\\
 & What are the consequences of coral reef degradation?\\
\bottomrule
\end{tabular}
\end{table}

\begin{table}[ht!]
\caption{Challenges and opportunities in performing causality-based study for Earth science.}
\label{tab:chlng_opp}
\begin{tabular}{lll}
\toprule
\textbf{Area} & \textbf{Challenges} & \textbf{Opportunities} \\ 
\midrule
   Causal Discovery &  Accurate modeling of complex processes that & Domain Knowledge based Causal Discovery\\ 
   & occur in the Earth's system  & \\
   Causal Discovery & Lack of ground truth information for validation& Synthetic datasets simulating Earth science data \\
   &  of causal models & distributions\\
   Causal Discovery & No information on true causal frequency & Multi-scale causal discovery \\ 
   Causal Discovery / Causal Inference & Tackling biases in simulated data & Adjusting for hidden confounders\\
   Causal Discovery / Causal Inference & High dimensional data & Extracting causally relevant variables, \\
   & & Causal representation learning, Causal discovery \\
   & & in low dimensional space\\ 
   Causal Discovery / Causal Inference & Non-linear data distributions & Deep learning based Causal Discovery / \\ 
   & & Inference methods  \\ 
   Causal Discovery & Seasonality / Non-stationarity in data & Incorporating periodicity / segmentation,\\
   & & Sliding window analysis \\ 
   Causal Inference & Tackling bias due to time-variying confounding & Matching Methods Treatment weighting methods \\ 
   Causal Inference & Tackling bias due to spatial confounding & Neighboringhood weighting, Adjusting \\
   & & spillover effects\\ \bottomrule
\end{tabular}
\end{table}

\section{Open Challenges in Causality-based Earth science Study}
Due to the multi-way interactions between the atmosphere, hydrosphere, cryosphere, geosphere, and biosphere, studying causality between them is a challenging but important task. 
Here, we have identified several fundamental questions within the Earth science domain, particularly concerning the rapidly increasing emissions of greenhouse gases and the consequential impacts of climate change, which may benefit from causality methods. These inquiries pose significant challenges due to several factors:
 1) Data availability may be inadequate or insufficient, hindering comprehensive analysis.
 2) Climate data is inherently complex, with multiple dimensions, making analysis and interpretation challenging. 
3)  The complex interaction among numerous variables complicates the disentanglement of causal relationships and feedback mechanisms. 
4) Confounding effects further hinder the clarification of causality. 
5)  The inability to conduct controlled laboratory experiments in climate science limits our understanding of causal relationships. 
6) The current state of climate models may lack the sophistication and precision necessary to accurately capture the complex interactions among variables. In Table~\ref{tab:earth_chlng}, we enlist some of the causality-related open questions in different Earth science domains \citep{masson2021climate, portner2022ipcc}. In Table \ref{tab:chlng_opp}, we identify some of the challenges in performing observational causal discovery and causal inference in various domains of Earth science and the opportunities they pave for researchers. 
\section{Causal Discovery}
Causal discovery refers to the process of discovering causal relations and uncovering underlying data generation process in real world data, helping researchers make more informed decisions, predict outcomes, and understand how changes in one variable, state or process may lead to changes in others \citep{pearl:88}.
In recent years, several graph-based algorithms are proposed for discovering the causal structures from empirical data \citep{spirtes2000causation,tian2013causal}.  The important causal connections are discovered purely based on the statistical analyses of observation data. This section presents an overview of time-series and spatiotemporal causal discovery, the key concepts in causality, and common approaches for performing causal structure learning for both temporal and spatial datasets. It further includes brief overview of renowned causal discovery methods, their limitations and applications in Earth science study.
To begin with, we have enlisted the key terminologies in causal discovery in Table~ \ref{tab:key-terms}.
\begin{table}[!ht]
\centering
\caption{Key terminologies in Causal Discovery}
\label{tab:key-terms}
\begin{tabular}{ll}
\toprule
\textbf{Terminology} & \textbf{Explanation} \\ \midrule
Causal discovery models & Discovers causal relationships between variables\\ 
Undirected Graph & A graph with no direction between the nodes \\ 
Probabilistic Graphical Model & Captures conditional independence relationships between random variables. \\ 
Bayesian Network &   Uses probability theory and a Directed Acyclic Graph (DAG) to quantify the relationships between variables. \\ 
Latent parameter  & Unknown / hidden parameters  \\ 
Conditional independence & Nodes that are not causally related \\ 
Conditional dependence & Nodes that are causally related \\ 
Markov Equivalence Class & A set of Directed Acyclic Graphs (DAGs) that may vary in structure (with different node arrangements and arrow directions) but all represent the same set of conditional independence relationships among the variables. \\
Correlation graph & It determines the correlation patterns in the data \\ 
Structural Causal Model & It provides a mathematical and graphical representation of causal relationships \\ 
Chain & Series-like structure (Figure~\ref{fig:chain}: X1 causes X2 and X2 causes X3) \\ 
Fork & This structure shows a common cause of two effects. (Figure~\ref{fig:chain}: X2 causes X1 and X3)\\ 
Collider & This structure include two causes of a single effect. (Figure~\ref{fig:chain}: both X1 and X3 are the causes of X2) \\ 
Partial correlation & Partial correlation refers to lagging each of several variables on all relevant delays relative to another variable  \\ 
Mutual independence & Each event is independent of any combination of other events in the collection  \\ 
Structure learning & The process of inferring the structure of directed acyclic graph (DAG) from data  \\ 
Moralization & Cognitive enhancement in causal selection process \\ 
Probability & Likelihood occurrence of an event \\ 
\bottomrule
\end{tabular}
\end{table}
\begin{figure}[!htbp]
\centering
\includegraphics[width=0.6\linewidth]{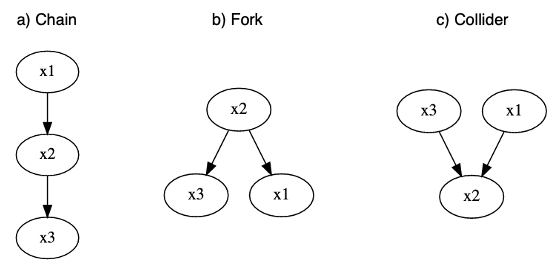}
  \caption{Three structures of causal graphical model. The nodes represents entities and edges represent direction of causal relation.}\label{fig:chain}
\end{figure}

\subsection{Causal Assumptions}

There are several foundational principles and conditions known as causal assumptions that are essential for inferring causal relationships between variables. Various causal discovery methods rely on different sets of these assumptions. Below are some commonly recognized causal assumptions:

\paragraph{Causal Sufficiency:} 

\begin{figure}[!htb]
        \center{\includegraphics[height=3cm, width=3cm]
        {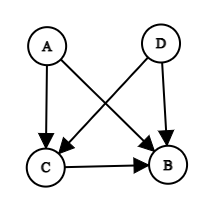}}
        \caption{\label{sufficiency} An example to demonstrate causal sufficiency.}
    \end{figure}

According to this condition, a pair of variables is considered causally sufficient if all their common causes are measured, meaning there are no latent confounders \citep{nogueira2021causal}. Causal sufficiency is a very stringent assumption because the concept of a closed world does not exist. In real-world scenarios, it is often the case that not all possible causes are measured, leading to frequent violations of this assumption \citep{pellet2008finding}. Causal insufficiency occurs when, for a set of variables \( V \), there are variables not included in \( V \) that are direct causes of more than one variable within \( V \). In Figure \ref{sufficiency}, the variables \( A, C, B \) are causally insufficient because another variable \( D \) outside of set \( V \) is the common cause of \( B \) and \( C \).

    \paragraph{Causal Markov Condition (CMC) / Causal Markov property:} 
    
\begin{figure}[!htb]
        \center{\includegraphics[height=3cm, width=3.5cm]
        {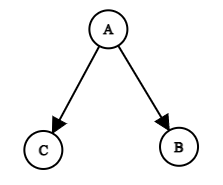}}
        \caption{\label{Markov} An example to demonstrate the causal Markov condition/property.}
    \end{figure}

    The principle behind the CMC is that all pertinent probabilistic information about a variable can be derived from its direct causes (parents), even if the variable’s descendants (effects) are ignored \citep{scheines1997introduction}. The CMC can also be understood through d-separation. In a causal DAG \( G \), variables that are d-separated will be conditionally independent in the corresponding probability distribution \citep{weinberger2018faithfulness}. This implies that among three disjoint nodes, one node must block all paths between the other two nodes \citep{nogueira2021causal}. The causal Markov assumption holds in a DAG that is causally sufficient, meaning there are no latent confounders. The CMC for the variables in the causal graph of Figure \ref{Markov} is as follows:\\
    B $\perp$ C $|$  A (B and C are independent of each other conditioned on their common parent A)\\

    \paragraph{Causal Faithfulness Condition (CFC):} 
    

    The causal faithfulness assumption can be described as follows: In a causal DAG \( G \), no conditional independence between variables exists unless it is implied by the causal Markov condition. Essentially, CFC identifies which variables in a DAG will be probabilistically dependent \citep{weinberger2018faithfulness}. This is the opposite of the Markov condition, which specifies which variables will be conditionally independent given their parents. In terms of d-separation, CFC indicates that except for the d-separated variables in a DAG, all other variables are dependent. In Figure~\ref{Markov}, the variables \( (A, C) \) and \( (A, B) \) are dependent. According to CFC, this implies that no observed interdependency in the data is accidental, but rather results from the underlying causal mechanism \citep{druzdzel2009role}.

    \paragraph{Acyclicity:} 
    
\begin{figure}[!htb]
        \center{\includegraphics[height=3cm, width=3.5cm] 
        {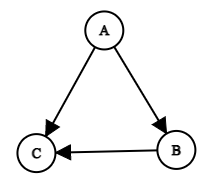}  \hspace{2cm} \includegraphics[height=3cm, width=3.5cm] {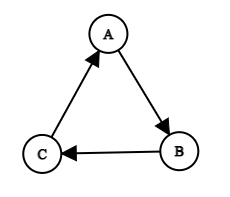}
        
        (a)\hspace{5cm}(b)}
        \caption{\label{ACYC} An example to demonstrate the acyclicity property where the graph in (a) holds acyclicity and (b) violates acyclicity.}
    \end{figure}

    The acyclicity assumption asserts that no feedback loops are permitted in a directed causal graph. This means there cannot be a directed path from a variable leading back to itself. This is a fundamental property of a causal graph, as shown in Figure~\ref{ACYC}, which must be satisfied to accurately represent causal relationships. Therefore, a causal graph is often represented using a DAG where no variable is an ancestor or descendant of itself \citep{greenland2007causal}. Adhering to the acyclicity condition is essential for the proper application of many causal inference techniques, as it prevents the complications that arise from circular causality and helps maintain the integrity of the causal model.

    \paragraph{Data Assumptions:}

    Causal discovery approaches significantly depend on the available data and the assumptions made about it. The data used for causal analysis can be observational, interventional, or both. To perform causal discovery, various assumptions about the data distribution are often considered. These assumptions include whether the data is linear or nonlinear, continuously valued or discrete valued, and whether it is stationary or non-stationary (for temporal data). Additionally, assumptions about different noise distributions, such as Gaussian, Exponential, or Gumbel noise, to which the data may belong, are commonly made. Other frequent assumptions about the data include the presence of any sampling or selection bias, missing data, latent variables, or noise. These assumptions are crucial because they provide the necessary framework for accurately inferring causal relationships from data. Without these assumptions, the causal inferences drawn could be misleading or incorrect, ultimately compromising the conclusions and applications derived from the causal analysis.

\subsection{Evaluation Metrics}
Evaluating the performance of causal discovery methods is crucial to assess their accuracy and reliability in identifying causal relationships from observational data. Given below are the commonly used evaluation metrics for causal discovery:
    \paragraph{True Positive Rate (TPR)} is the likelihood ratio of true causal edges to be identified correctly by the causal method. Assuming the threshold $t$ of the probability of an edge $Prob(e_{ij})\in (0,1)$, the calculation of TPR is: 
    \begin{equation}
      TPR_t = |\frac {\{ (i,j):Prob(e_{ij})\geq t \}\cap S} {S}|  
    \end{equation} 
    where $S$ is the set of edges whose ground truth is positive.

    \paragraph{False Positive Rate (FPR)} is the likelihood ratio of absence of causal edges to be identified incorrectly by the causal method. Assuming the threshold $t$ of the probability of an edge $Prob(e_{ij})\in (0,1)$, $FPR$ is defined as
    \begin{equation}
    FPR_t = |\frac {\{ (i,j):Prob(e_{ij})\geq t \}\cap \hat S} {\hat S}|
    \end{equation}
    where $\hat S$ is the set of edges that do not exist in the true causal graph.
    
    \paragraph{Structural Hamming Distance (SHD)} represents number of operators required to make two partial DAGs match. The operators include adding or removing an undirected edge, and adding, removing or reversing the direction of an edge. The algorithm of SHD can be found in\citep{tsamardinos2006max}.

    
    \paragraph{Structural Intervention Distance (SID)} measures number of vertex pairs $(i,j)$ in the estimated Directed Acyclic Graph (DAG) that correctly predict the intervention distributions among the distributions that are Markov equivalent with respect to another DAG.

    \paragraph{Area Over Curve (AOC)} simplifies area under curve (AUC) with $1-AUC$. The curve is plotted with $TPR_t$ against $FPR_t$ as $(FPR_t, TPR_t)$, where $t \in [0,1]$.

\subsection{Common Approaches}
Causal discovery techniques are typically divided into three main types or approaches:
\paragraph{Constraint based Approach:}

This is a widely used method in causal discovery, which identifies the causal relationships by testing for conditional independence among variables. Within this framework, the Peter-Clark (PC) algorithm \citep{spirtes2000causation} is a foundational technique. It operates under the assumption of faithfulness, meaning all dependencies in a Directed Acyclic Graph (DAG) must comply with the d-separation principle. Another prominent method in this category is the Fast Causal Inference (FCI) algorithm \citep{spirtes2000causation}. Unlike the PC algorithm, FCI does not assume causal sufficiency, acknowledging the possibility of unobserved confounders in the analysis of real-world data.

\paragraph{Score based Approach:}

The score-based approach is a prevalent method in causal discovery that seeks to identify the causal graph that best fits the data according to a given score function, such as the Bayesian Information Criterion (BIC). This approach requires to search across all potential graphs to maximize the score, reflecting the graph's fit to the data \citep{triantafillou2016score}. Due to the nature of the approach, it can be computationally intensive, especially as the number of variables increases, leading to an exponential growth in the number of possible graphs \citep{nogueira2021causal}. A notable algorithm within this category is the Greedy Equivalence Search (GES) algorithm \citep{chickering2002optimal}, which begins with an empty graph and progressively adds or removes edges based on their contribution to the overall score, optimizing the graph structure step-by-step.

\paragraph{Functional Causal Models-based Approach:}

Functional Causal Model (FCM) approaches, also referred to as SCM-based methods, define causal relationships using specific functional forms, such as a set of equations where the variables are functions of their causal parents plus an independent noise term ($y = f(pa_{y}) + e$). 
FCM-based techniques aim to differentiate between different DAGs within the same Markov equivalence class by making assumptions about the underlying data generation functions. A diverse array of FCM-based approaches exist for both temporal and i.i.d. data, which address linear and non-linear causal relationships through methods such as Independent Component Analysis (ICA) \citep{stone2004independent} and Additive Noise Models (ANMs) \citep{hoyer2008nonlinear} respectively.

\subsection{Time-Series Causal Discovery}
Time-series causal discovery extends the causal structure learning process to temporal data. Causal discovery in time-series data involves dealing with the specific challenges of temporal dependencies and potential feedback loops. It has been widely used in econometrics, neuroscience, and climate science \citep{spirtes2000causation, zhou2014analysis, van2015causal, ghysels2016testing}. Identifying time lags is an important aspect of causal discovery in time-series data. It involves determining the delay or time difference between the cause and its effect. Time lags are crucial for understanding the dynamics of the causal relationship and for making accurate predictions in time-series forecasting tasks. To identify time lags in causal relationships, researchers may use methods like cross-correlation analysis, autocorrelation, or time-series cross-validation techniques \citep{spirtes2000causation}. These methods help measure the strength of the association between variables at different time lags and assist in determining the optimal lag that maximizes the predictive performance or the strength of the causal relationship. We explain some of the widely used causal discovery methods below, along with their variants, underlying causal assumptions and limitations. A tabulated summary of these time-series causal discovery methods is given in Table~\ref{tab:CD-Methods-Comparison-Table}.

\begin{table}[!h]
\centering
\caption{Comparison among the causal discovery methods for time series data. Here (\xmark ) is mentioned against the methods where the assumptions or relation is non-existent or not explicitly mentioned in the paper.}
\label{tab:CD-Methods-Comparison-Table}
\begin{tabular}{c|ccc|cc|cc}
\toprule
\multirow{2}{*}{Method} & \multicolumn{3}{c|}{Assumptions} & \multicolumn{2}{c|}{Causal Relationship} & \multicolumn{2}{c}{Produced Edge} \\ \cline{2-8} 
 & \multicolumn{1}{c|}{Faithfulness} & \multicolumn{1}{c|}{Sufficiency} & \begin{tabular}[c]{@{}c@{}}Causal Markov \\ Property\end{tabular} & \multicolumn{1}{c|}{Linear} & Non-linear & \multicolumn{1}{c|}{Instantaneous} & Lagged \\ \midrule
Granger Causality & \multicolumn{1}{c}{\xmark} & \multicolumn{1}{c}{$\checkmark$} &  \xmark& \multicolumn{1}{c}{$\checkmark$} &  \xmark& \multicolumn{1}{c}{\xmark} &  \xmark\\
PCMCI & \multicolumn{1}{c}{$\checkmark$} & \multicolumn{1}{c}{$\checkmark$} & $\checkmark$ & \multicolumn{1}{c}{$\checkmark$} & $\checkmark$ & \multicolumn{1}{c}{\xmark} & $\checkmark$ \\ 
PCMCI+ & \multicolumn{1}{c}{$\checkmark$} & \multicolumn{1}{c}{$\checkmark$} & $\checkmark$ & \multicolumn{1}{c}{$\checkmark$} & $\checkmark$ & \multicolumn{1}{c}{$\checkmark$} & $\checkmark$ \\ 
LPCMCI & \multicolumn{1}{c}{$\checkmark$} & \multicolumn{1}{c}{$\checkmark$} & $\checkmark$ & \multicolumn{1}{c}{$\checkmark$} & $\checkmark$ & \multicolumn{1}{c}{\xmark} & $\checkmark$ \\ 
LiNGAM & \multicolumn{1}{c}{\xmark} & \multicolumn{1}{c}{$\checkmark$} &  \xmark& \multicolumn{1}{c}{$\checkmark$} &  \xmark& \multicolumn{1}{c}{\xmark} &  \xmark\\ 
VarLiNGAM & \multicolumn{1}{c}{\xmark} & \multicolumn{1}{c}{$\checkmark$} &  \xmark& \multicolumn{1}{c}{$\checkmark$} &  \xmark& \multicolumn{1}{c}{$\checkmark$} & $\checkmark$ \\ 
TS-LiNGAM & \multicolumn{1}{c}{\xmark} & \multicolumn{1}{c}{$\checkmark$} &  \xmark& \multicolumn{1}{c}{\xmark} &  \xmark& \multicolumn{1}{c}{\xmark} &  \xmark\\ 
TiMINo & \multicolumn{1}{c}{\xmark} & \multicolumn{1}{c}{\xmark} &  \xmark& \multicolumn{1}{c}{\xmark} & $\checkmark$ & \multicolumn{1}{c}{\xmark} &  \xmark\\ 
NAVAR & \multicolumn{1}{c}{\xmark} & \multicolumn{1}{c}{\xmark} &  \xmark& \multicolumn{1}{c}{\xmark} & $\checkmark$ & \multicolumn{1}{c}{\xmark} & $\checkmark$ \\ \bottomrule
\end{tabular}
\end{table}

\subsubsection{Granger Causality, V-Granger}
Granger causality (GC) is a statistical hypothesis test to determine the significance one time series in predicting another \citet{granger1969investigating}. It is named after Clive Granger, a Nobel laureate economist who developed the concept. 
A time-series X "granger-causes" time-series Y if historic values of X play a significant role in predicting future values of Y. Granger causality is a statistical test and does not guarantee a true causal relation between two time-series. Nevertheless, Granger causality is commonly used in various fields, including economics, finance, engineering, and neuroscience, to explore relationships between time-varying data. However, it has limitations and assumptions. For instance, the choice of time lags can influence the results. GC assumes causal sufficiency condition, stationary and linear relationship and does not predict the direction of causal relation. Further, the method is data hungry and requires sufficiently large sample size. 

While Granger Causality is widely used in various fields to understand the causal relationship between time series, it can yield misleading results in the presence of unobserved confounders in  observational data. 
V-Granger \citep{meng2019estimating}, overcomes the causal sufficiency limitation by estimating Granger causality in the presence of unobserved confounders using a variational autoencoder (VAE). V-Granger incorporates an inference network to estimate the intractable posterior distribution while its generative network models the non-linear causal relationship in time-series data.  
Through experiments on synthetic and semi-synthetic datasets, as well as two real datasets: Butter and Temperature, the authors demonstrated how V-Granger outperforms the traditional Granger test in some cases. However, the method is not suitable for datasets with high noise levels. 

\subsubsection{PCMCI, PCMCI+, LPCMCI}
PCMCI\textsuperscript{\ref{note1}} \citep{runge2019detecting} is a method for estimating causal graphs from multivariate time series datasets. In addition to causal structure learning, PCMCI can also estimate direct and  mediated causal effects. A baseline of PCMCI was first introduced by \citep{runge2015identifying} for identifying causal pathways in spatiotemporal climate systems. The method combines the PC stable causal discovery algorithm with momentary conditional independence (MCI) in a two-step approach. The first step is to find lagged parents of all individual nodes using the PC stable method. The second step is to test the momentary conditional independence. As a result of the second step, if two nodes are independent, then there exists no cause-effect relationship between them. The outcome of this two-step method includes a causal adjacency matrix, corresponding lag values and strength of identified relationships given by p-values. PCMCI supports multiple linear and non-parametric conditional independence tests depending on the data distribution. For instance, the partial correlation or $ParCorr()$ test assumes linear dependencies and Gaussian noise in the data. There are multiple variants of PCMCI available for different downstream tasks. For instance, PCMCI+ extends the method for  auto-correlated nonlinear time series data to discovers instantaneous and lagged causal relations. For details on PCMCI's implementation, independence tests available and different variants of PCMCI, please refer to their Github\footnote{\label{note1}\url{https://github.com/jakobrunge/tigramite}} repository.  

Similarly, Latent PCMCI or LPCMCI\textsuperscript{\ref{note1}} \citep{LPCMCI} is a constraint-based approach which is a variant of the PCMCI algorithm for causal discovery from observational time series data. This method is suitable when the data is non-linear, comprises hidden confounders, and the goal is to learn lagged and contemporaneous causal relations. 

\subsubsection{LinGAM, VARLiNGAM, TS-LiNGAM }
LinGAM\footnote{\url{https://lingam.readthedocs.io/en/latest/index.html}} (Linear Non-Gaussian Acyclic Model) \citep{shimizu2006linear} is a framework used for causal discovery and estimation of causal effects from observational data. It focuses on identifying linear causal relationships among variables assuming non-Gaussian noise distributions and acyclic causal relations. VARLiNGAM \citet{hyvarinen2010estimation} extends the foundational LiNGAM model to encompass scenarios involving time series. This augmentation involves integrating the core LiNGAM model with the conventional vector autoregressive models (VAR). Consequently, it facilitates the exploration of causal relationships encompassing both delayed and concurrent (instantaneous) effects. This stands in contrast to the traditional VAR framework, which solely examines causal relations with a time delay.
VARLiNGAM holds similar causal assumptions as the base LinGAM model, such as, data linearity, causal sufficiency with no hidden confounders, acyclicity of contemporaneous causal relations and non-gaussian continuous error variables. 

TS-LiNGAM \citep{tsLiNGAM} is an extension of the LiNGAM algorithm for estimating both instantaneous and lagged causal edges in time-series data. The model is a combination of autoregressive models and structural equation modeling (SEM). It assumes the external noises to be mutually independent, temporally uncorrelated, and to be non-Gaussian. The overall method has four steps: i) Estimation of a classic autoregressive model, ii) Residuals computation, iii) Performing LiNGAM analysis on the residuals, and iv) Computation of the causal effect estimates. The method has been evaluated on real datasets such as financial (stock) data and magnetoencephalographic data.

\subsubsection{TiMiNo}

    TiMINo \citep{peters2013causal} approach provides a new method at that time for causal inference on time series data that overcomes some of the methodological issues of previous methods that could not capture nonlinear and instantaneous effects. Specifically, TiMINo extends Structural Equation Models (SEMs) for time series analysis, utilizing a restricted function class to achieve general identifiability results. This approach accounts for both lagged and instantaneous effects, which can be nonlinear and unfaithful, as well as non-instantaneous feedback within the time series. Given a sample of finite and continuous-distributed multivariate time series data, and output the summary time graph if the input data satisfies model assumptions. If the input does not meet the assumptions, TiMINo typically refrains from making causal determinations, avoiding incorrect causal conclusions. The model is represented as $X_t^i = f_i((\mathbf{PA_p^i})_{t-p}, \ldots, (\mathbf{PA_1^i})_{t-1},  (\mathbf{PA_0^i})_{t}, N_t^i)$, where $p$ is the time lag for independence test, $(\mathbf{PA_1^i})$ is the set of parent nodes for node $i$, $X_t$ is time-series data with finite and continuous distribution, and $N_t^i$ is a noise vector jointly independent over i and t. The prior assumption of this approach is that each time sequence can be represented by a function of all its direct causes and some noise variable, and restricts the function class to provide general identifiability results. Potential weaknesses of TiMINo are discussed including the possibility of fitting a model in the wrong direction and the effects of independence testing on smaller datasets.
    
\subsubsection{Tidybench algorithms}
    
\citep{weichwald2020causal} presents four algorithms\footnote{\url{https://github.com/sweichwald/tidybench}} for causal discovery from time series data to address some common challenges such as time aggregation, subsampling, and delays. The algorithms also focus on the properties of realistic weather and climate data and were among the winning algorithms in the Causality 4 Climate (C4C) competition at the NeurIPS 2019 conference. The output from these algorithms is a matrix that contains a score for each edge between a pair of variables. For example, for factors ($X_{i}$, $X_{j}$), the score represents the likelihood that the causal edge $X_{i}$ → $X_{j}$ exists. The algorithms are grounded in the following concept: present values are regressed on past values, and the regression coefficients are analyzed to determine if one variable Granger-causes another. Here is a brief overview of each algorithm: (i) SLARAC (Subsampled Linear Auto-Regression Absolute Coefficients) fits a VAR model to bootstrapped samples of the data by randomly choosing a lag to include in the model, (ii) QRBS (Quantiles of Ridge-regressed Bootstrap Samples) uses bootstrapped samples similar to SLARAC, but applies Ridge regression to correlate time delays with previous values, (iii) LASAR (Lasso Auto-Regression) conducts Lasso regression on the residuals obtained from the previous step against values further back in time, keeping track of which variables are selected at each lag, and eventually fitting an OLS regression using only the selected variables, and lastly, (iv) SELVAR (Selective auto-regressive model) employs a hill-climbing technique based on the leave-one-out residual sum of squares to determine edges. SELVAR scores these edges using their regression coefficients. 
\subsubsection{Additive Nonlinear Time Series Causal Models} Chu et al. \citep{chu2008search} introduced a method for learning additive non-linear time series, and demonstrated its effectiveness in extracting detailed causal information from stationary models. Before this, no previous methods existed for non-linear systems. Based on assumptions of stationary data, the methods discuss and extend linear methods including PC and Fast Causal Inference (FCI) with joint Normal distributions. The primary algorithm employed is PC, which is known to be consistent only when there are no feedback relations or latent common causes. The paper also mentions other related algorithms such as FCI and an algorithm by Richardson and Spirtes that allow for latent variables and linear feedback relations, respectively. However, no algorithm consistently identifies linear causal models in the presence of latent variables and feedback. To extend the PC and similar algorithms to a broader range of systems, including nonlinear continuous models, a more comprehensive conditional independence test is necessary. Therefore the paper proposes to use a new definition of the independence test used for the additive nonlinear time series model and embed it into a causal inference algorithm. The simulation results demonstrate the effectiveness of the additive non-linear algorithm in inferring causal structure from time series data, particularly for trigonometric lag models and linear lag models.

\subsubsection{Additional Methods}
There are several other causal discovery methods introduced for time-series data in recent years. Temporal Causal Discovery Framework (TCDF\footnote{\url{https://github.com/M-Nauta/TCDF}}) \citep{nauta2019causal} aims to uncover causal graph structures within observational time-series data by utilizing attention-based convolutional neural networks. This model processes multivariate time-series data and outputs a graph that includes nodes and edges with corresponding lagged values. TCDF employs the Permutation Importance Validation Method (PIVM) to validate potential causes. In the paper, TCDF's performance has been evaluated using financial stock return data and fMRI data for tracking brain blood flow.

Similarly, Neural Additive Vector Autoregression Models (NAVAR\footnote{\url{https://github.com/bartbussmann/NAVAR}}) discover highly non-linear causal relationships between factors for time series data. NAVAR employs deep neural networks to derive the additive Granger causal influences from the temporal dynamics of multivariate time series data. As the model name implies, NAVAR proposes an additive structure to extend VAR. The new method makes predictions linearly depend on independent nonlinear functions of the input variables, modeled through a multi-layer perceptron. This approach enables the scoring and ranking of causal relationships. NAVAR has been validated using various benchmark datasets for causal discovery available on the CauseMe\footnote{https://causeme.uv.es/} platform, particularly in the fields of Earth science.

TS-CausalNN \citep{faruque2024ts} learns causal graphs with lagged and instantaneous causal links simultaneously from non-stationary non-linear multivariate time series data. TS-CausalNN utilizes the concept of learning non-stationary features using non-linear functions and proposes a novel parallel neural network design using a custom causal convolution 2D layer. The model processes temporal data without any external adjustments like detrending, deseasonalizing, etc., and generates a full causal graph with link values. Three synthetic datasets and three real-world datasets with non-stationary and non-linear features were used to assess the effectiveness of this approach.

Another causal discovery method based on deep learning is Causal-HMM\footnote{\url{https://github.com/LilJing/causal_hmm}} \citep{li2021causal}. The method proposes to predict irreversible diseases at an early stage using time series data. It is an HMM-based hidden Markov model consisting of a subset of hidden variables, and a reformulated sequential variational autoencoder (VAE) framework to learn the proposed causal hidden Markov model. The model is applied to early prediction of peripapillary atrophy on an in-house dataset of peripapillary atrophy (PPA).

\subsection{Spatiotemporal Causal Discovery}
Spatiotemporal causal discovery refers to the process of identifying causal relationships among variables that vary both in space and time. This field typically deals with datasets where observations are collected over both spatial and temporal dimensions, such as data from weather systems, ecological systems, social networks, or biological processes. The goal of spatiotemporal causal discovery is to uncover the underlying causal structure that governs the dynamics of these systems. This involves not only identifying causal relationships between variables but also understanding how these relationships change over space and time. Here we present some of the recent methods proposed to perform causal discovery on spatiotemporal datasets.


    \subsubsection{Mapped-PCMCI}
\begin{figure}[!htb]
    \centering
    \includegraphics{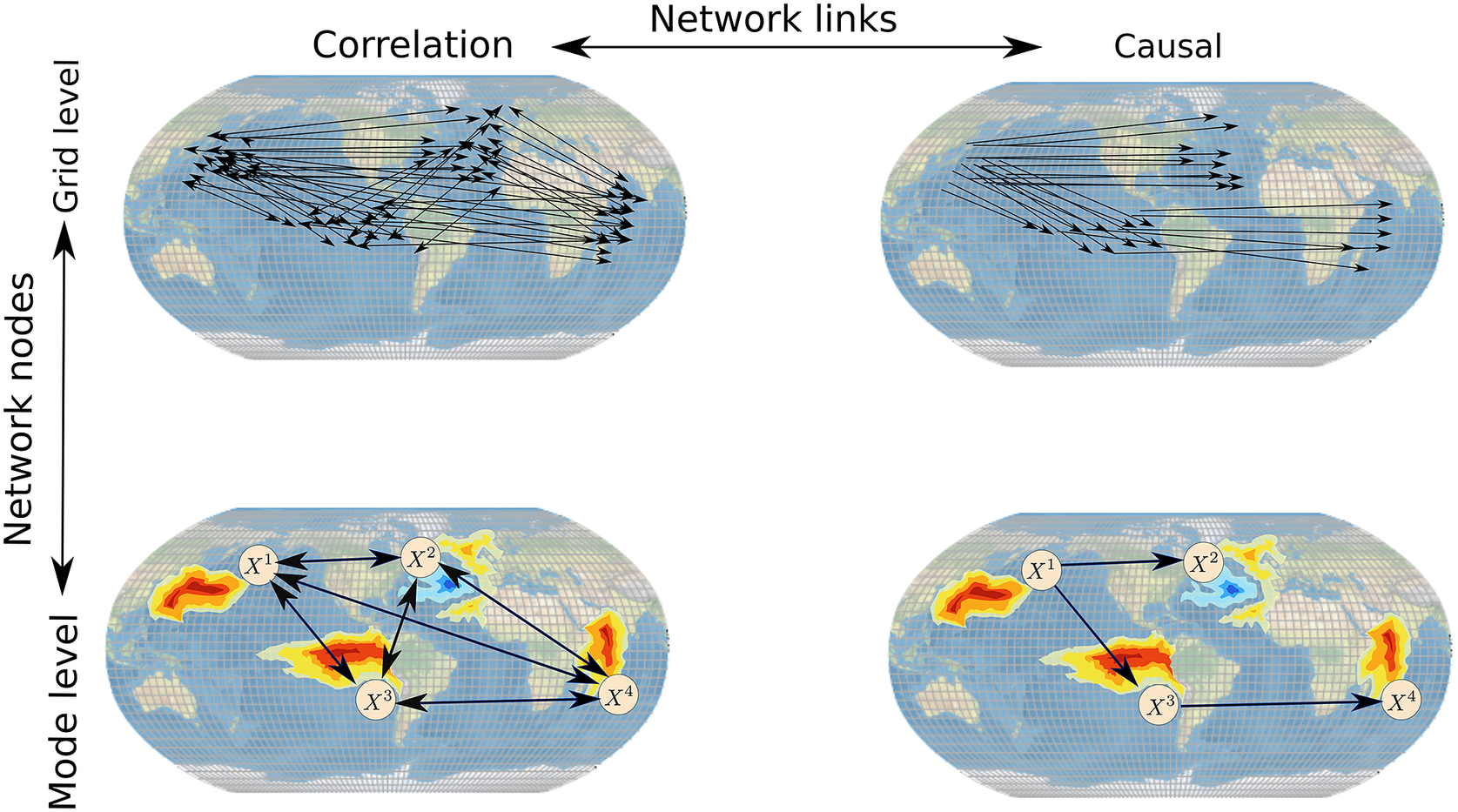}
    \caption{An application of spatiotemporal causal discovery method (Mapped-PCMCI) for identifying teleconnections. \citep{tibau2022spatiotemporal}}
    \label{fig:mapped-PCMCI}
\end{figure}

    \citep{tibau2022spatiotemporal} presented a spatiotemporal stochastic climate model SAVAR which can be used to benchmark causal discovery methods for teleconnections, providing insights into the strengths and weaknesses of different analysis methods. The authors also introduced novel causal discovery method named Mapped-PCMCI contributing to causality-based climate model evaluation. 
    Mapped-PCMCI is based on the assumption that the causal dependencies within a gridded dataset can have a lower-dimensional latent representation. The method consists of four steps (Figure~\ref{fig:mapped-PCMCI}): (i) Dimensionality reduction method on the gridded data, (ii) Application of causal discovery method to the lower-dimensional time series variables, (iii) Estimation of (lagged) causal effects, (iv) Inverse transformation of lower-dimension features back to the original grid locations. This method aims to overcome the challenges of dealing with large networks, nonstationary networks, and the high-dimensional causal estimation of spatiotemporal data. It provides a spatial grid-level, allowing for the analysis of causal relationships in complex systems. Python code for both the SAVAR and Mapped-PCMCI is accessible on GitHub \footnote{\url{https://github.com/xtibau/}}.

    \subsubsection{Interactive Causal Structure Discovery (ICSD)} 
    

    \cite{melkas2021interactive} introduced a method called Interactive Causal Structure Discovery (ICSD) designed to integrate expert knowledge during the application of causal discovery algorithms in Earth science. ICSD allows users to interactively modify initial algorithm outputs, enabling the iterative inclusion of expert knowledge into initial models. This method treats structure learning as a task of creating a probabilistic model based on the data, incorporating expert knowledge as prior distributions across possible causal structures. The current method uses a greedy approach, which can get stuck in local optima due to the dependence on the initial model state. Each causal model is evaluated based on a score reflecting its log-likelihood. For practical illustration, ICSD was applied to a forestry dataset with variables like shortwave downward radiation and temperature using three causal discovery algorithms: PC, GES, and LiNGAM. The findings indicate that even minimal prior knowledge can enhance the performance of these algorithms, and highlighted the potential for overfitting and concept drift, which ICSD can identify and address. Currently, the system starts with outputs from various causal discovery algorithms, which is not ideal. Future work aims to identify optimal starting points for user exploration.

    \subsubsection{Spatio-Temporal Causal Discovery Framework(STCD)}

    The Spatio-Temporal Causal Discovery Framework (STCD) \citep{STCD} aims to identify causal relationships in spatiotemporal data by enforcing both temporal and spatial constraints. 
    STCD primarily extends the TCDF \citep{TCDF} framework by adding a component that ensures the enforcement of the spatial constraint. Using the TCDF attention scores, it filters the list of potential candidates. The proposed framework STCD eliminates irrelevant candidates by penalizing the attention scores if the candidates violate the spatial constraints.  
    This framework takes influence from hydrological systems, where a river's geographic location gives it temporal precedence over other rivers, based on the direction of water flow from one river to the other. Motivated by this idea, the spatial constraint imposes a direction of the flow. Specifically, it is  the product of a distance $d$ and a spatial coefficient $\lambda$. The distance has either a positive or negative direction based on the geographical positioning of the locations, and $\lambda$ controls the effect of the spatial constraint on the attention score. A limitation of STCD is that it cannot model the spatial and temporal interactions simultaneously, which might hinder the discovery process when the spatial and temporal components influence each other.
    
\subsubsection{Group Elastic Net}

\citep{lozano2009spatial} proposed a data-centric approach to climate change attribution, using spatial-temporal causal modeling to analyze climate observations and forcing factors. The authors developed a novel method called Group Elastic Net to infer causality from the data and incorporate extreme value modeling to study extreme climate events. With the assumption of spatial stationarity, the model combines graphical modeling techniques with Granger Causality for spatiotemporal data. The model also applies extreme-value theory to model extreme events and incorporates these estimates into the causal modeling and attribution process. The experiments involve data collection from multiple sources. For real-world data collection, the researchers obtained data from various sources including the Climate Research Unit (CRU), the National Oceanic and Atmospheric Administration (NOAA), NASA, the National Climate Data Center (NCDC), and the Carbon Dioxide Information Analysis Center (CDIAC). These sources provided data on climate variables such as temperature, precipitation, solar irradiance, greenhouse gases, and aerosols. In addition to the real climate data, the researchers used a spatial-temporal vector autoregressive (VAR) model to generate synthetic data. The experiments compared the performance of the Group Elastic Net method, which considers spatial interactions, with a method that neglects such interactions. Utilizing the method, the researchers estimated the influence of CO2 and greenhouse gases on increasing temperatures. Their proposed approach can offer a useful alternative to simulation-based climate modeling.

\subsubsection{pg-Causality}

Zhu et al. \citep{zhu2017pg} presented a novel approach, called pg-Causality, to efficiently identify spatiotemporal causal pathways for air pollutants using urban big data by combining pattern mining and Bayesian learning. The approach overcomes the challenges of noise, computational complexity, and complex causal pathways, and thus outperforming traditional methods in terms of time efficiency, inference accuracy, and interpretability. The authors use the FEP Mining Algorithm to mine frequent episode patterns (FEPs) in a symbolic pollution database. This algorithm considers constraints such as consecutive symbols being different and a specified temporal constraint between consecutive records. After discovering the FEPs, the authors extract candidate causes for each sensor by finding pattern-matched pairs within a specified time lag threshold. The authors use a Gaussian Bayesian network (GBN) based graphical model to identify causal pathways among air pollutants. The authors refine the causal structures learnt from GBN using an expectation-maximization (EM) learning phase and a structure reconstruction phase. 
The experiments were conducted using real-world dataset of 6 air pollutants and 5 meteorological measurements from from North China, Yangtze River Delta, and Pearl River Delta. 
The experimental results demonstrated pg-Causality's high accuracy, time efficiency, and scalability in inferring causal relationships in air pollution data.

\subsection{Applications of Causal Discovery in Earth science }
We explained some of the renowned causal discovery methods for time-series and spatiotemporal data in the previous subsection. Here, we will share some of the applications of those widely used causal discovery methods, specifically in the Earth science domain. A summary of these applications is provided in Table~\ref{tab:cd-app}.

\begin{table}[]
\caption{Applications of time-series and spatiotemporal causal discovery methods in Earth science applications.}
\label{tab:cd-app}
\begin{tabular}{lll}
\toprule
\textbf{Data} & \textbf{Method}         & \textbf{Application}                                                                           \\
\midrule
\multirow{10}{*}{Time-series}   & PC                   & Causal discovery for hydrometeorological systems (Ombadi et al. 2020)                                           \\
              & GC                      & Long-term causal links in climate change events (Smirnov and Mokhov 2009, Kodra et al. 2011)   \\
              & GC                      & Causal discovery for hydrometeorological systems (Ombadi et al. 2020)                          \\
              & GC                      & Causal discovery for teleconnections (Mosedale et al. 2006, Varando et al. 2021)               \\
              & CEN                     & Causal discovery for midlatitude winter circulation within the Arctic (Kretschmer et al. 2016) \\
              & CEN                     & Causal discovery for precursors of september Arctic sea-ice extent (Li et al. 2018)            \\
              & PCMCI                   & Causal discovery for biosphere–atmosphere interactions (Krich et al. 2020)                     \\
              & PCMCI                   & Causal discovery to study wildfire impact (Qu et al. 2021)                                     \\
              & Causal Feature Learning & Causal discovery for teleconnections (Chalupka et al. 2016)                                    \\
                                & Tidybench Algorithms & Causal discovery for time-aggregation, time-delays and time-subsampling in \\
                                & & weather data (Weichwald et al. 2020) \\ \hline
\multirow{6}{*}{Spatiotemporal} & PC-stable            & Spatiotemporal causal discovery for univariate climate data (Ebert-Uphoff and Deng 2017)                        \\
              & PCMCI                   & Causal discovery for tropical–extratropical summer interactions (Di Capua et al. 2022)         \\
              & Mapped-PCMCI            & Causal discovery for teleconnections (Tibau et al. 2022)                                       \\
              & STCD                    & Causal discovery for hydrological systems (Sheth et al. 2022)                                  \\
              & Group Elastic Net       & Causal discovery for climate change attributions ((Lozano et al. 2009)                         \\
              & pg-Causality            & Identifying causal pathways for air pollutants (Zhu et al. 2017)   \\
              \bottomrule
\end{tabular}%
\end{table}

    \subsubsection{Applications of Granger Causality}
     
      In climate research, understanding complex phenomena, such as teleconnection patterns, is important because it links atmospheric changes in one region to impacts in distant regions. However, the automatic identification of these patterns from observational data is still unresolved due to nonlinearities, nonstationarities, and the limitations of correlation analyses. Varando et al.\citep{varando2021learning} propose a deep learning approach to address these problems and learn Granger causal feature representations that capture the true causal effects of the target index, such as El Niño Southern Oscillation (ENSO) or North Atlantic Oscillation (NAO). The authors propose a method called the Granger Penalized Autoencoder with the assumptions including the presence of nonlinearities and nonstationarities in the observational data, as well as the limitation of correlation analyses in identifying true causal patterns.  By utilizing normalized difference vegetation index (NDVI) data collected from MODIS reﬂectance data over 11 years, their work identified clear patterns of the causal footprints of ENSO on vegetation in different regions. The GitHub repository of this research is approachable\footnote{\url{https://github.com/IPL-UV/LatentGranger}}.
      \citep{mosedale2006granger} used a Granger causality based approach to quantitatively measure the feedback of daily sea surface temperatures (SSTs) on daily values of the North Atlantic Oscillation (NAO). This was done by simulating a realistic coupled general circulation model (GCM). This study is an extension of the work by Mosedale et al. in 2005. where the Granger causality approach is used to find the best time series models for modeling the coupled system for greater flexibility. \citep{smirnov2009granger} introduced the idea of long-term Granger causality to find out how strongly the global surface temperature (GST) is affected by variations in carbon dioxide, atmospheric content, solar activity, and volcanic activity during the last 150 years. \citep{kodra2011exploring} extended the classic Granger causality test and hypothesized the multisource causal relation between globally averaged land surface temperature (GT) observations and observed CO2 in the atmosphere.

     \subsubsection{Applications of Pearl Causality}

     Causal discovery algorithms based on probabilistic graphical models have been applied in geoscience applications to identify and visualize dynamical processes, but the lack of ground truth and unexplained connections have posed several challenges. To address these challenges, Ebert-Uphoff et al.\citep{ebert2017causal} developed a simulation framework using synthetic spatiotemporal data to better understand the physical processes and interpret the resulting connectivity graphs, ultimately solving the mystery of the previously unexplained connections. This approach allows for the resolution of previously unexplained connections and provides a benchmark for other causal discovery algorithms. The authors used a constraint-based structure learning method called the PC stable algorithm, which is a modification of the classic PC algorithm. The PC stable algorithm has advantages such as increased robustness of results and suitability for parallelization. The authors dropped the requirement of causal sufficiency and focused on necessary conditions for cause-effect relationships. The article discusses the results of three different scenarios in the simulations. In Scenario 1, increasing the spatial resolution leads to more edges for a specific time interval and fewer edges for a longer time interval. In Scenario 2, concurrent edges are believed to be caused by contradictory velocities at the boundaries of the advection field. In Scenario 3, concurrent edges in the center do not match the typical diffusion pattern and appear to fill modeling gaps. The velocity estimates in Scenario 1 with higher speed are weak due to many high-speed interactions represented as concurrent edges.

     \subsubsection{Applications of PCMCI}
     \citep{PCMCI-application-1} used the PCMCI algorithm to study the underlying causal relations in biosphere–atmosphere interactions. Particularly they estimated the causal graphs from the eddy covariance measurements of land–atmosphere fluxes and global satellite remote sensing of vegetation greenness datasets. The causal graphs revealed the gradual shifts that correspond to little adjustments, such as the relationship between temperature and visible heat as well as increasing dryness which might not have been discovered merely through correlation analysis. 
     
     \citep{PCMCI-app-3} used PCMCI to recover the causal graphs for 8 vegetation types which represent the causal relations and time lags between wildfire burned areas and weather/drought and vegetation conditions. A significant conclusion they found is that weather and aridity conditions are dominant indicators to burned areas for grassland. Also, for broadleaf forests, radiation while for needleleaf forests temperature is the most vital indicator. To analyze the influence of a set of spatial patterns representing tropical–extratropical summer interactions, \citep{PCMCI-app-2} estimated causal maps which is an extension of PCMCI to spatial fields of variables. 

     Lastly, given the challenges in analyzing teleconnections and the lack of ground truth benchmark datasets, Mapped-PCMCI\citep{tibau2022spatiotemporal} proposed to tackle these challenges by presenting a simplified stochastic climate model that generated gridded data and represents climate modes and their teleconnections.

     \subsubsection{Applications of other Causal Discovery Methods}
\begin{figure}[!htbp]
\centering
\includegraphics[width=0.7\linewidth]{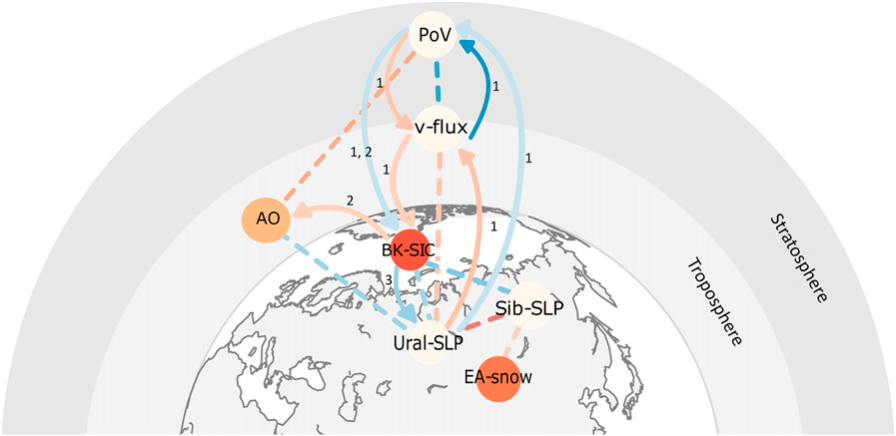}
  \caption{Causal graph discovered using Causal Effect Network (CEN) identifying regional actors of Arctic winter circulation based on their monthly mean time-series data. \citep{kretschmer2016using}}\label{fig:cen}
\end{figure}

     One of the predecessors of PCMCI method is the Causal Effect Network \citep{kretschmer2016using} which was implemented to identify the time-delayed causal relationships between different actors of midlatitude winter circulation within the Arctic. Through experiments on monthly, bi-monthly and quarter-monthly time-series of seven meteorological variables (See \citep{kretschmer2016using} for details), the authors pin-pointed Barents and Kara sea ice to be important drivers of winter circulation further confirming the troposphere-stratosphere coupling proposed in previous literature. The causal graph of this discovery is given in Figure~\ref{fig:cen}. The same approach was utilized by \citep{li2018precursors} to study the precursors of summer (September) Arctic sea-ice extent. 

     \citep{Chalupka2016elnino} learned macro-level causal features from micro-level variables without supervision using Causal Feature Learning (CFL). These macro-level features are used to discover causal relationships between them and target output. Aggregating micro-level variables into the macro-level variables helps to ignore changes in the micro-level variables that have no effect on the output variable Y. Here the authors calculated conditional expectation  $E[Y|x]$ using the regression of Y based on x. Then these conditional expectations are clustered to generate the macro-level variables. This method was used to find the causal relation between the El Nino and La Nina events with the eastern Pacific near-surface wind (ZW, zonal wind) and sea surface temperature (SST) variable.    

     \citep{ombadi2020evaluation} aimed to evaluate different causal discovery methods and their performance in retrieving causal information from synthetic data and real-world observations for hydrometeorological systems. The methods for causal discovery in hydrometeorological systems involved in the paper are Granger causality (GC), Transfer Entropy (TE), graph-based algorithms (PC), and Convergent Cross Mapping (CCM). The author applies them to examine the causal drivers of evapotranspiration in a shrubland region during summer and winter seasons. The study aims to present the fundamentals of these methods and shed light on their assumptions in the context of hydrometeorological systems. Specifically, it discusses the assumptions of causal sufficiency, causal faithfulness, and stationary time series. To evaluate performance of the four causal discovery methods, the study uses synthetic data generated from the bucket model and analyzes the causal structures of hydrological systems. The evaluation includes assessing the asymptotic performance of each method, investigating the sensitivity to sample size, and assessing the sensitivity to the presence of noise. The observational data is obtained from the Santa Rita Mesquite FluxNet site in Arizona to analyze the environmental controls on evapotranspiration.

\section{Causal Inference}
Causal inference or causal effect estimation is the process of quantifying the causal impact of an entity (state or event) that contributes to the data-generation process of another entity (outcome, state or event). 
Causal inference has been applied to study environmental science for several decades, with early applications dating back to the mid-20th century \citep{hill1965environment}. However, much of the earlier causal inference based analysis was done on independent and identically distributed (i.i.d) data utilizing statistical inferential and regression techniques to estimate the causal effects on potential outcomes \citep{pearl2009causal}. This section presents the key terminologies (Table~\ref{tab:key-terms-ci}), common approaches and causal assumptions required to perform causal inference techniques. We explain the different time-series and spatiotemporal causal inference techniques introduced, along with their limitations and applications in Earth science. 

\begin{table}[ht!]
\centering
\caption{Key terminologies in causal inference}
\label{tab:key-terms-ci}
\begin{tabular}{lll}
\toprule
\textbf{Terminology} & \textbf{Explanation} \\ \midrule
causal effect & strength or infleunce of a causal relation \\ 
instance & a single unit; data sample \\ 
treatment & cause; variable that is intervened on \\ 
potential outcome & effect; variable exposed to the treatment \\ 
confounders  &  variable influencing both treatment and outcome \\ 
covariates  & pre-treatment variables; features \\ 
intervention & nudging the value of a treatment variable \\ 
average treatment effect (ATE) & the average difference in potential outcomes with and without\\
& undergoing intervention\\

\bottomrule
\end{tabular}
\end{table}

\subsection{Common Approaches}
Causal inference methods are based on one of the two main categories: do-calculus and potential outcome framework. 

\subsubsection{Do-Calculus}
 The first method, do-calculus was developed in 1995 to identify causal effects in non-parametric models using conditional probabilities \citep{spirtes2010introduction}. Once a causal structure is identified, do-calculus can be applied to find interventional distributions by deriving mathematical representations for a physical intervention using the $do()$ operator, as shown in Equation \ref{eq:docalc}. Here, Y represents the outcome, X represents the variable intervened on and Z represents a set of covariates.
 \begin{equation}
    \begin{multlined}
     P(Y | do(X = x)) = \Sigma_{Z} P(Y = y | X = x, Z = z)\\
     P(Z = z)
     \end{multlined}
     \label{eq:docalc}
 \end{equation}
 In case of a binary-valued variable X, the average causal effect (ACE) can be calculated using do-calculus by calculating the difference between do(X=1) and do(X=0), as shown in Equation \ref{eq:acedocalc}.
  \begin{equation}
     ACE = P(Y | do(X = 1)) - P(Y | do(X = 0))
     \label{eq:acedocalc}
 \end{equation}
 
\subsubsection{Potential Outcome Framework:}
The potential outcome framework, also known as the Rubin Causal Model (RCM), is a formal approach to causal inference in statistics and social sciences \citep{rubin2005causal}. It provides a way to define and estimate causal effects by considering hypothetical scenarios, also known as treatments. For each unit (e.g., individual, event, etc.) in the study, the potential outcome framework considers the outcomes under each possible treatment. If there are two treatments, $T=1$ and $T=0$, then each unit can have one of two potential outcomes: $Y(1)$: the outcome of the unit after undergoing treatment, and $Y(0)$: the outcome of the unit without receiving treatment. For any unit, we can only observe one of the potential outcomes (the one corresponding to the treatment actually received). The other potential outcome is counterfactual and unobservable. This makes direct calculation of causal effects impossible without further assumptions or methods.

\subsection{Assumptions for Causal Inference}
For consistent causal effect estimation on observational data, it is important to hold the following identifiability conditions or causal assumptions:
\paragraph{Consistency:} Under the consistency condition, the potential outcome for the treated subject $Y_{X=1}$ is considered equivalent to the observed outcome Y. The same goes for the untreated subject.
\paragraph{Positivity:} This assumption implies that the probability of receiving treatment given some covariates $Z$ is always greater than zero. That is, $Pr(X = x | Z = z) > 0$ where $Pr(Z =z) \neq 0$. 
\paragraph{Conditional Exchangeability:} Under the conditional exchangeability assumption, also known as "weak ignorability", the condtional probability of receiving treatment depends only on the covariates $Z$, that is, $Y_x$ and treatment $X$ are statistically independent given every possible value of $Z$. On the contrary, unconditional exchangeability implies that treatment group has the same distribution of outcomes as the untreated control group.
\paragraph{SUTVA:} Under the Stable Unit Treatment Value Assumption (SUTVA), the potential outcome $Y_i$ on one unit $i$ is not affected by the treatment effect on other units and there is no hidden variations of treatment.

\subsection{Evaluation Metrics}
\subsubsection{Root Mean Square Error (RMSE)}
The RMSE score is a widely used evaluation metric to calculate error in regression-based predictions. In case of causal inference, this metric is used for evaluating the performance of predictive models when ground truth information about factual and counterfactual data is known. RMSE error can be computed using the formula in Equation ~\ref{eq:rmse} where $Y_i$ represents the actual value of the outcome and $\hat{Y_i}$ represents the estimated value of the $i^{th}$ potential outcome by a causal inference method.
\begin{equation}
    RMSE = \sqrt{\frac{1}{N} \Sigma_{i=1}^N(Y_i - \hat{Y_i})^2}
    \label{eq:rmse}
\end{equation}

\subsubsection{Precision in Estimated Heterogeneous Effects (PEHE)}
PEHE is a metric similar to the RMSE score that is used for calculating the error in the actual and predicted average treatment effects (ATEs) \citep{hill2011bayesian}. Just like RMSE score, PEHE scores are only applicable when counterfactual values are already known, a case common in synthetic datasets.
\begin{equation}
    \sqrt{PEHE} = \sqrt{\frac{1}{N} \Sigma_{i=1}^N(ATE_i - \hat{ATE_i})^2}
\end{equation}

\subsection{Time-series Causal Inference}
Real world observations require dynamic or time-varying causal analysis focusing on calculating the impact of an intervention on a sequential or time-varying outcome \citep{moraffah2021causal}. For instance, policymakers and climate change activists would be interested in identifying the impact of lowering CO2 emissions on the rate of ozone depletion over a specific period of time. Time series causal inference provides a framework for understanding and quantifying causal relationships in data that evolve over time, employing a range of statistical and econometric techniques to address the unique challenges posed by temporal data such as non-stationarity and time-varying confoundedness. A non-stationary time series is a time series whose statistical properties, such as mean, variance, and autocorrelation, change over time. Non-stationarity exhibits spurious correlations in data that leads to incorrect causal effect estimations. Confoundedness also comes as a significant challenge because it can lead to incorrect conclusions about the causal relationship. Confounders are variables (or covariates) whose past values influence the future values of both the cause (treatment) and the effect (outcomes). When confounding is not properly accounted for, the observed association between the cause and effect may be due to the confounding variable rather than a true causal effect leading to biased predictions. Both non-stationarity and confounding are prevailing challenges in causal inference for Earth science applications. 

We summarize some of the widely-used causal inference techniques for time-varying and time-invariant data, and their applications in Earth science, below. 
\subsection{Time-varying Causal Inference Methods}
Time-varying causal inference methods are approaches used to understand and analyze causal relationships in situations where the treatment or intervention, the outcome, and potentially the covariates, change over time. These methods aim to uncover how a changing treatment influences the outcome of interest, considering the dynamic nature of both the treatment and the outcome variables.

In contrast to traditional causal inference, where the focus is on a fixed intervention and its effect, time-varying causal inference takes into account the evolving nature of interventions and outcomes. This is particularly relevant in fields such as epidemiology, economics, and environmental science, where interventions and exposures can vary over time. Some common time-varying causal inference methods include:
\subsubsection{Marginal Structural Models}
Marginal Structural Models (MSMs) \citep{robins2000marginal} comprise methods designed to estimate the effects of time-varying treatments while tackling the challenge of time-varying confoundedness. Since traditional methods like ordinary regression models may not properly handle time-varying treatments and confounders, MSMs were developed to address biases that can arise when using these traditional regression models to estimate causal effects in situations where treatments change over time. MSMs provide a framework to model and adjust for the dynamic nature of treatments, confounders, and their interdependencies using the IPTW weights. MSMs first employ IPTW to re-weight the data and emulate a hypothetical time-fixed treatment scenario. This mitigates confounding by making treated and untreated groups comparable. A weighted regression model is then employed to estimate the treatment effect, accounting for the dynamic confounding. MSMs rely on the assumptions of positivity and no unmeasured confounders. Though MSMs have become a cornerstone in time-varying causal inference in the fields of epidemiology \citep{vanderweele2009marginal, hernan2000marginal}, public health \citep{williamson2017marginal, vanderweele2011marginal} and social sciences\citep{bacak2015marginal}, they also have some limitations. Complex interactions between time-varying treatments and confounders, longitudinal missing data and information censoring can be challenging to model using MSM technique. 

\subsubsection{Convergent Cross Mapping}
Convergent Cross Mapping (CCM) 
\citep{ye2015distinguishing} is a nonlinear causal inference technique used to detect the presence of causal relationships in time-varying data. It is particularly useful when the relationship between variables is complex and non-linear. CCM is based on the concept of time delay embedding, where time series data is transformed into higher-dimensional space by embedding time-delayed copies of the series.

\subsubsection{Instrument Variables}
In causal graphs, a variable is called an instrumental variable (IV) if it is independent of the hidden confounders and related to the effect only through the cause. For a causal model of the form $Y = \beta X+g(H,\varepsilon^{Y})$ prediction of the Y based on the observation X with the presence of hidden confounders yields a biased estimation of the $\beta$ coefficient. The Conditional Instrumental Variable (CIV) \citep{CIV-2022} method used IVs to identify the $\beta$ coefficient from the time series causal model with the presence of hidden confounders applying condition on the required number of previous instances of IVs.  


\subsubsection{Deep Representation Learning based Models}
In recent years, there has been growing interest in leveraging the powerful representation learning capabilities of deep learning models to tackle challenges in causal inference \citep{bengio2013representation, koch2021deep}. By encoding complex relationships in the data, deep learning based CI methods can learn a lower-dimensional representation of covariates, which can be used to predict counterfactual outcomes, treatment effect estimation and even control for confounding. Neural networks can also be used to estimate propensity scores, which represent the probability of treatment assignment given covariates. These propensity scores can then be used for matching, weighting, or as covariates in outcome models. For time series data, deep learning models can be used to handle the temporal dependencies and estimate causal effects over time \citep{moraffah2021causal}. Recurrent Marginal Structural Networks (R-MSN) \citep{lim2018forecasting}, Time-series deconfounder \citep{bica2020time} and Counterfactual Recurrent Network (CRN) \citep{bica2020estimating} are some of the deep learning based causal inference techniques that deploy g-methods to tackle time-varying confounding, however these techniques can only work for binary treatment effect estimation. Whereas, TCINet \citep{ali2023quantifying} and G-Net \citep{li2021g} can work for both binary and continuous treatment effect estimation, making them more suitable for Earth science domains. 


\subsection{Time-invariant Causal Inference Methods}
Time-invariant treatment effect estimation methods focus on quantifying the impact of an intervention on the outcome when both the intervened and un-intervened outcome variables are observed at the same time. Whereas, the effect of time-invariant intervention is measured based on the difference in the outcomes before and after the intervention takes place. We enlist some of the time-invariant causal inference methods below:
\subsubsection{Causal-ARIMA}
The Causal-ARIMA (C-ARIMA) is a potential outcome framework-based method to estimate causal effects from time series datasets \citep{causal-arima}. This method is suitable for datasets where it is not possible to separate the treated and controlled groups to estimate the effects of the applied treatment. In an observational study, if an intervention event occurs at a specific time point, the time series divides into two different time horizons, pre-intervention and post-intervention. Here the intervention is not applied at each time point, it occurred at a given time and draws the effect on the rest of the observations of the system. The C-ARIMA method is a three-step process: first, the relationship between the covariates and target variable is learned from the pre-intervention data using an ARIMA model. Then using learned relationships from pre-intervention data the counterfactual outcomes are predicted for post-intervention data points. Finally, the causal effect is measured from the factual observations and predicted counterfactual data. The single persistent intervention, non-anticipating prospective outcomes, temporal no-interference, independence of covariates and treatment, and unforeseen treatment are the five underlying principles that underpin the C-ARIMA method's operation. The C-ARIMA method can identify different types of causal effects without making any structural assumptions about the effect caused due to the intervention. 

\subsubsection{Difference in Difference}
The Difference-in-Differences (DID) method \citep{did2011} is a statistical technique used in econometrics and social sciences, that considers a treatment group and a control group to assess the causal influence of a treatment, intervention, or policy by comparing changes in outcomes between two groups over time. DID is particularly useful when randomization of subjects into treatment and control groups is not possible or ethical, making it challenging to establish causality through traditional experimental methods. To perform causal inference using this method, start by selecting two groups: a treatment group that is exposed to the intervention or policy change and a control group that is not exposed. These groups should be similar in all relevant aspects except for the treatment or policy change. Next, data is collected on the outcome of interest for both groups, typically before and after the treatment or policy change. Finally, the causal effect is estimated by calculating the difference in the changes in outcomes in the controlled group and the treated groups. The DID estimate can be analyzed using regression techniques to control for potential confounding factors, such as demographics or other trends affecting the outcome. However, there are some limitations. The method requires adequate sample size to yield meaningful results and it can handle time-varying confounders. Overall, Difference-in-Differences is a powerful tool for causal inference when randomized controlled trials are not feasible or ethical. However, it relies heavily on the validity of the parallel trends assumption, and researchers must carefully consider potential sources of bias when designing and interpreting DID studies. The mathematical formula to quantify the effect of the applied treatment is given as:
\begin{equation}
     \text{DiD Estimate} = (Y_T(\hat{X}) - Y_T(X)) - (Y_C(\hat{X}) - Y_C(X))
     \label{eq:did}
 \end{equation}
Here, $Y_T$ refers to the outcome in treated group, $Y_C$ refers to the outcome in the controlled group, $X$ refers to the pre-treatment data and $\hat{X}$ refers to post-treatment data.
\subsubsection{Interrupted Time Series}
Interrupted time series (ITS) analysis \citep{mcdowall1980interrupted} is a statistical method used for assessing the causal impact of an intervention, treatment, or policy change on a specific outcome by analyzing changes in a time series data set. Unlike the Difference-in-Differences (DID) method, which compares two groups (treatment and control) before and after an intervention, ITS focuses on a single group and tracks changes in that group's outcomes over time, before and after the intervention. ITS requires defining a point in time when the intervention, treatment, or policy change occurred. This is referred to as the "interruption point". A statistical model is built to estimate the expected change direction of the outcome in the absence of the intervention. Common models used for pre-intervention estimation include linear regression, autoregressive integrated moving average (ARIMA), or segmented regression models. The intervention effect is defined as the variation between the expected and observed results at each time point following the intervention.

Interrupted time series analysis is particularly valuable for evaluating the impact of policies, interventions, or treatments that are implemented at a specific point in time. It provides a rigorous framework for establishing causal relationships by demonstrating that observed changes in the outcome are more likely due to the intervention than to other factors. However, as with any statistical method, the validity of ITS analysis depends on the quality of data and the appropriate modeling of the pre-intervention trend. ITS can be a powerful tool for causal inference when experimental designs are not feasible. 

\subsubsection{Causal Impact}
The Causal Impact method \citep{causalimpact2015} is a statistical approach for temporal data to estimate the causal effect of an intervention, event, or treatment. It was developed by Google's Research and Data Science teams and is implemented in the R package \footnote{\url{http://google.github.io/CausalImpact/CausalImpact.html}}. This method is particularly useful for businesses and researchers to assess the impact of marketing campaigns, policy changes, or any other interventions on metrics such as sales, website traffic, or user engagement. The Causal Impact method builds a Bayesian structural time series (BSTS) model to represent the expected behavior of the time series data in the absence of the intervention. The BSTS model captures trends, seasonality, and other patterns in the pre-intervention data. The difference between the observed post-intervention data and the counterfactual prediction represents the causal impact of the intervention. This difference is used to estimate the treatment effect and assess its statistical significance. Causal Impact is particularly valuable when you want to assess the impact of a specific event or intervention on a time series metric while accounting for the underlying patterns and variations in the data. It offers a probabilistic and data-driven approach to causal inference, which can be beneficial for decision-making in various domains.

\subsubsection{Regression Discontinuity Design}
The Regression Discontinuity Design (RDD) method \citep{hahn2001identification} is a quasi-experimental research design used in causal inference to estimate the causal effect of a treatment, intervention, or policy change when the assignment to the treatment or control group is determined by a cutoff or threshold in a continuous or ordinal variable. RDD is particularly useful when random assignment to treatment and control groups is not feasible, but there exists a natural "discontinuity" in the assignment process that can be exploited to make causal inferences. RDD is a powerful method for causal inference when the assignment to treatment or control groups is determined by a discontinuity in a continuous or ordinal variable. However, it relies on the assumption of a smooth discontinuity and requires careful design and analysis to ensure valid causal inferences.

RDD estimates the treatment effect using regression analysis. The method requires defining a threshold $c$ in the data $X$. The ATE is estimated by comparing the average outcomes for individuals just below the threshold $(X < c)$ with those just above the threshold $(X >= c)$. The difference in outcomes between the two groups can be attributed to the treatment or intervention. Mathematically, the ATE estimate can be expressed as follows:
\begin{equation}
     ATE = E(Y|X >= c) - E(Y|X < c)
     \label{eq:rdd}
 \end{equation}

\begin{table}[!htbp]
\centering
\caption{Comparison of existing causal inference methods for temporal data and their relevance to Earth science.}
\label{tab:tsci-methods}
\begin{tabular}{@{}cccccc@{}}
\toprule
Method &
  \begin{tabular}[c]{@{}c@{}}Binary/\\ fixed treatment\end{tabular} &
  \begin{tabular}[c]{@{}c@{}}Continuous \\ treatment\end{tabular} &
  \begin{tabular}[c]{@{}c@{}}Time varying \\ treatment\end{tabular} &
  \begin{tabular}[c]{@{}c@{}}Time varying\\  covariates\end{tabular} &
  \begin{tabular}[c]{@{}c@{}}Applicable on \\ Earth science \end{tabular} \\ \midrule
Difference in Difference &  \cmark   &    \xmark &  \xmark   & \xmark    &   \xmark  \\ 
Causal Impact & \cmark  & \cmark & \cmark & \xmark &   \cmark  \\ 
Instrument Variables     &  \cmark   &   \cmark  &   \cmark  &   \cmark  &   \cmark  \\ 
CRN  &   \cmark  &  \xmark   &   \cmark  &  \cmark   &   \xmark  \\ 
MSM & \cmark & \xmark  & \cmark & \xmark  &   \xmark  \\ 
R-MSN  &   \cmark   &  \xmark    &  \cmark   &  \cmark   &   \xmark\\ 
Time-series Deconfounder &  \cmark   &  \xmark    &  \cmark   &  \cmark   &   \xmark  \\ 
TCINet &  \cmark   &  \cmark    &  \cmark   &  \cmark   &   \cmark  \\ 
G-Net &  \cmark   &  \cmark    &  \cmark   &  \cmark   &   \cmark  \\ \bottomrule
\end{tabular}
\end{table}

\subsection{Spatiotemporal Causal Inference}
Methods for spatiotemporal causal inference often integrate techniques from both causal inference and spatiotemporal analysis. These methods may include statistical modeling, machine learning algorithms, and deep learning approaches to estimate causal effects in observational data while considering spatial and temporal dependencies. Reich et al.  provided a comprehensive review of the challenges and limitations of performing causal inference in spatial settings and enlisted methods that exploit spatial structure and those that account for spatial interference in spatiotemporal settings \citep{reich2021review}. Here, we present some of the recent methods introduced to perform causal inference on spatiotemporal data.
\subsubsection{Propensity Score Based Spatiotemporal Causal Inference Methods}

Propensity scoring techniques are extended in spatiotemporal causal inference setting to account for confounding caused by spatial confounders. Papadogeorgou et. al. proposed a distance adjusted propensity score method to incorporate spatial proximity in causal inference techniques in the presence of observed confounders \citep{papadogeorgou2019adjusting}. Later, they proposed a spatiotemporal causal inference method for stochastic interventions assuming both the treatment and outcome variables are coming from the spatiotemporal point processes \citep{ST-causal-22}. In the stochastic intervention setting the treatments are specified through a probability distribution instead of fixed values in the potential outcome framework. 
This model makes two assumptions on the treatment: unconfoundedness and bounded relative overlap of the treatment, which ensures that the intervention with high probability does not become a fixed treatment. This method allows the intervention can be applied to any spatial location, an infinite number of interventions at the same time and measures the outcome for all applied interventions using an intervention distribution pattern. So it is possible to apply and measure the outcomes for any spatial and temporal intervention patterns.
\begin{figure}
    \centering
    \includegraphics{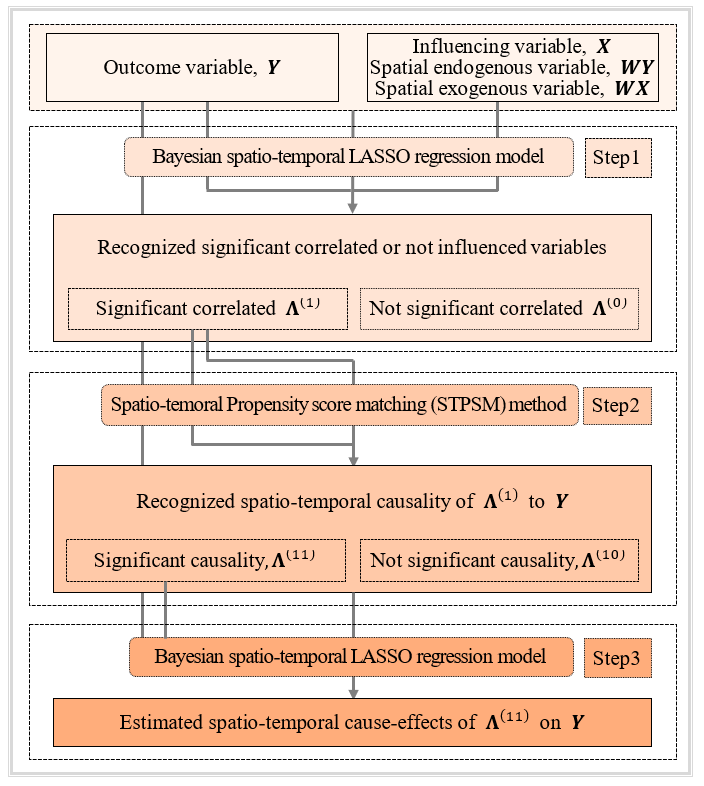}
    \caption{A schematic flow-diagram of spatiotemporal causal inference method proposed by \citep{li2023insighting}}
    \label{fig:stci}
\end{figure}
\citep{li2023insighting} proposed a method for recognizing spatiotemporal causality and effect estimation, integrating a spatiotemporal propensity score matching (STPSM) and Bayesian spatiotemporal LASSO regression model (BST-LASSO-RM). This three-step method is illustrated in Figure~\ref{fig:stci}. 

\subsubsection{Spatially Interrupted Time-Series (SITS) }
\citep{zhang2023spatiotemporal} investigated the spatiotemporal heterogeneities in the causal effects of mobility intervention policies during the COVID-19 outbreak using a spatially interrupted time-series (SITS) analysis. The 
\citep{zhang2023spatiotemporal} conducted a study to understand the spatiotemporal heterogeneity and the causal effectiveness of mobility intervention policies during COVID-19 outbreak, particularly at finer-grained spatial and temporal resolutions. Their research provided insights for policymakers on when and where to implement specific policies to control the spread of the virus. The SITS system involves the application of the Interrupted Time-Series (ITS) design combined with spatial analysis. The ITS design is a quasi-experimental design for policy evaluation that records and measures behavioral outcomes before and after implementing a policy intervention. The SITS analysis requires a long time series with multiple observations over time, especially before the treatment period, to establish a credible estimation of counterfactuals. It compares mobility outcomes over time within a single group of individuals or neighborhoods, rather than between a treatment group and a control group. This helps to control for potential biases arising from between-group differences. To incorporate spatial heterogeneities, the SITS analysis considers the spatial contexts of neighborhoods, including socioeconomic and built environment settings. The effects of the policy intervention, such as the level and slope changes, are moderated by these spatial contexts. The SITS model estimates the spatiotemporally varying causal effects of the policy interventions by incorporating these spatial moderating effects. The SITS analysis in this study used a time-series dataset of neighborhood-based mobility levels extracted from a large set of mobile phone signaling data in Shenzhen, China. The dataset allowed for tracking the daily mobility trend at the neighborhood level before and after the COVID-19 outbreak. The study analyzed the effects of two mobility intervention policies: the first-level response (FLR) policy and the closed-off management (COM) policy.

\subsubsection{Spatio-Temporal Causal Inference Network (STCINet)}
Most recently, \citep{ali2024estimating} proposed a deep learning based causal inference technique by extending the potential outcome framework for spatiotemporal data. Their proposed method, STCINet, is a multi-input model with a U-Net based stream for counterfactual prediction and a latent factor model for handling time-varying confounders. STCINet works under the assumption of no unmeasured confounding and is able to estimate treatment effects of space-time-varying interventions. The goal is two handle two major challenges in performing spatiotemporal causal inference. One, to reduce bias caused by time-varying confounders. Two, to incorporate spatial interference into the outcome predictions. The authors evaluate their model on two diffusion based synthetic datasets, with and without spatial interference, while they demonstrate the performance of their method on real-world Arctic sea ice predictions by estimating the impact of atmospheric interventions on summer sea-ice forecasting in different sub-regions within the Arctic. 

\subsubsection{Additional Spatiotemporal Causal Inference Methods}
Christiansen et al. proposed a latent spatial confounder model to study relationship between conflict and the forest loss in Columbia \citep{christiansen2022toward}. Papadogeorgou et al. recently proposed a Bayesian modeling approach to causal inference in the presence of latent confounders and spatial interference \citep{papadogeorgou2023spatial}. 
Overall, spatiotemporal causal inference still remains a lesser tapped area with much focus on spatiotemporal causal structure learning (part of spatiotemporal causal discovery). One reason behind this is the challenging task of inferring causation in both space and time-varying setting in the presence of latent confounders and spatial interference, two of the major challenges in this research area. 

\subsection{Applications of Causal Inference in Earth Science}
In Earth science applications of causal inference are constrained by the inability to validate the findings as observational data cannot be controlled, randomized or intervened on. Nevertheless, we enlist some of the recent inroads in causal inference for Earth science domains. 
\citep{yang2022detecting} applied an advanced data-driven causal inference method based on the convergent cross mapping (CCM) algorithm that provides valuable insights to identify thermal influence and feedback among the different climate regions in the contiguous U.S using the long-term (1901–2018) near surface air temperature observations.
\citep{gao2022temporally} studied some widely applied causal inference approaches to study the extent of the influence of different vegetation structures on net primary production (NPP) from both temporal and spatial perspective. Granger Causality Test (GCT), CCM, etc. are some of the used temporal causal models in this study. The authors conclude that in scenarios where it is difficult to infer causation from a temporal perspective, it may be feasible to use spatial causal inference there.
\citep{wu2023assessing} used a CCM method to study the driving effects of various environmental factors such as air temperature, vegetation index, soil moisture, net surface radiation, precipitation, water vapor, etc. on land surface temperature in China based on the remote-sensing and reanalysis data from 2003–2018.

\section{Datasets and Toolbox for Causality}
Datasets play a vital role in validating and analyzing the effectiveness of any new model or method. The ground truth knowledge of the dataset helps to approve or reject the result produced through any model. In this section, we provide reference to multiple synthetic, simulation and real-world datasets that can be utilized in causal discovery and inference methods benchmarking in different science domains. A summary of the publicly available toolboxes for causal study is given in Table~\ref{tab:tools}. 

\begin{landscape}
\begin{table}
\caption{Synthetic, simulated, and real world datasets used for causal analysis.}
\label{tab:datasets}
\begin{tabular}{lllll}
\toprule
\textbf{Dataset} &
  \textbf{Description} &
  \textbf{Data} &
  \textbf{Causal} & 
  \textbf{Dataset} \\
  &
   &
  \textbf{Type} &
  \textbf{Category} & 
  \textbf{Type} \\
%
\midrule
Harvard Dataverse \footnote{\url{https://dataverse.harvard.edu/dataverse/basic_causal_structures_additive_noise}}  & Contains six synthetic datasets representing different causal structures. The time series datasets are generated using  & Time & Causal & Synthetic \\
     & a nonlinear function of cause variables, linear self-causation and additive Gaussian noise. & series & Discovery &\\
\hline
FLAIRS  & This resource contains 22 simulated time series datasets. All datasets contain 20 continuous variables and   & Time & Causal & Synthetic\\
 \citep{FLAIRS} &  1000 time points with the lag of 1 and 3 time units. & series & Discovery &\\
\hline
FinanceCPT &  This simulated financial dataset collection contains 20 datasets with 25 variables and 10 causal structures. So two  & Time & Causal & Synthetic\\
 \citep{kleinberg2013causality} & datasets for the same causal structure. These datasets have [20, 40] random relationships between variables with   & series & Discovery &\\
 & 1-3 time lags and 4000 time steps. & & &\\
\hline
PROMO   & This is a dataset of product sales and different promotions for each product. The daily sales value for each & Time & Causal &
  Synthetic \\
    \citep{promo}  & product is recorded for three years with ground truth graph where about 1000 promotions applied on   & series & Discovery & \\
   & 100 different products. & & & \\ 
\hline
Diffusion Data & This dataset contains 4000 samples of diffusion-based spatiotemporal images. The dataset contains 3 variables  & Spatio- & Causal & Synthetic \\
\citep{ali2024estimating} & including treatment, time-varying confounder, and (factual and counterfactual) outcomes. & temporal & Inference & \\
\hline
North American Mesoscale & Generated by the National Centers for Environmental Prediction (NCEP) using the WRF Non-Hydrostatic Mesoscale  & Spatio- & Causal & Realistic\\ 
 (NAM)& Model. This is a spatiotemporal dataset of 12km resolution covering the continental United States and the data frequency  & temporal & Discovery &  Simulated \\
  \citep{nam_data}  & is every 6 hours from 2012-01-01 00:00 to 2023-10-15 18:00. Different properties of Air Temperature, Geopotential  & & &\\
  & Height, Humidity, Sea Level Pressure, Snow, Surface Pressure and Upper Level Winds are available in this simulation. & & &\\
\hline
FMRI benchmark & Functional Magnetic Resonance Imaging (FMRI) dataset contains 28 distinct simulations for various brain networks.   & Time & Causal & Realistic \\ 
 \citep{fmri} & Each of these causally sufficient dataset contains the neural activity based on the blood flow change and up to 50 time  & series & Discovery & Simulated \\
 & series with 50 and 5000 time steps.& & \\
 \hline
 DREAM4 & Is a simulated gene expression dataset based on patterns found in model organisms. This dataset contains time series and  & Time & Causal & Realistic\\
 \citep{dream4} & steady-state settings data for learning the gene regulatory network. Five distinct datasets are available in DREAM4, & series & Discovery & Simulated\\
 & each with ten distinct time-series recordings for one hundred genes spread over 21 time steps. & & &\\
\hline
NCEP-DOE Reanalysis 2 & The US National Centers for Environmental Protection (NCEP) and the Department of Energy (DOE) provide this  & Spatio- & Causal & Realistic\\ 
product  & dataset from 1979 to the current time. All available data is applied to a complex climate model to generate   & temporal & Discovery & Simulated \\
 \citep{ncep-doe} & reanalysis data for unobserved locations and missing time steps. This is a large set of almost 40 atmospheric variables  & & &\\
 &  measured in the reanalysis dataset. The dataset covers 90N-90S, 0E-357.5E  with a  & & &\\
 & 2.5-degree latitude x 2.5-degree longitude global grid (144x73). & & &\\

\bottomrule
\end{tabular}
\end{table}
\end{landscape}

\begin{landscape}
\begin{table}[ht]\ContinuedFloat
\caption{(Continued) Synthetic, simulated, and real world datasets used for causal analysis.}
\label{tab:datasets-2}
\begin{tabular}{lllll}
\toprule
\textbf{Dataset} &
  \textbf{Description} &
  \textbf{Data} &
  \textbf{Causal} & 
  \textbf{Dataset} \\
  &
   &
  \textbf{Type} &
  \textbf{Category} & 
  \textbf{Type} \\
%
\midrule
CausalWorld & This open-source causal structure learning benchmarking data generation platform contains a robotic environment & Time-series& Causal & Realistic \\  
  & manipulation dataset for different tasks. The generated datasets represent different causal structures of interacting  & & Discovery / & Simulated \\
 & objects like robot and object masses, colors, sizes, etc. Different causal studies like do-interventions, counterfactual & & Inference &\\
 & situations, structure learning, inference, etc. can be performed and evaluated using this platform. & &  & \\
 \hline
Beijing Multi-Site  &   This observational dataset was collected by the Beijing Municipal Environmental Monitoring Center and contains  & Time-series & Causal & Real-world \\
  Air-Quality Dataset & hourly observation of 6 pollutants in the air: CO, PM2.5, PM10, O3, NO2 and SO2, and 6 meteorological variables:  & & Discovery /  & \\
  \citep{beijing_multi-site_air-quality_data_501} & air temperature, wind direction and speed, pressure, dew point temperature, and precipitation. These data were  & & Inference & \\
  & collected from 2013 to 2017. & & \\
\hline
ERA5 & The European Centre for Medium-Range Weather Forecasts (ECMWF) maintains this global climate and weather & Spatio- & Causal & Real-world \\ 
  \citep{era5_data}  &  dataset. The hourly observations of the different atmosphere, land and oceanic variables are available & temporal & Discovery /  & \\
  &  in this dataset from 1940 to the present day for the whole globe and are updated daily for new data. & & Inference & \\
\hline
Sea Ice Data  & This data collection is a polar sea ice observational dataset maintained by the National Snow and Ice Data   &Spatio- & Causal & Real-world \\
 \citep{sea_ice} & Center (NSIDC). This dataset is collected from the Scanning Multichannel Microwave Radiometer & temporal & Discovery /  & \\
 &  (SMMR) instrument on the Nimbus-7 satellite and the Special Sensor Microwave/Imager (SSM/I). Several   & & Inference &\\
 & observational variables like sea ice concentration and extent, sea surface temperatures, wind stress, snow   & & &\\
 & cover, rainfall rates, etc. are recorded in this dataset from 1978 to the present. & & &\\

\hline
Metropolit Cohort & The Metropolit Cohort dataset contains data from 11532 humans born in 1953 and lived till 1968 in the Copenhagen  & Time-series & Causal & Real-world \\ 
\citep{Metropolit} & Metropolitan area, Denmark. This dataset comprises physical, medical, mental, social and diagnosis & & Discovery & \\ 
 & information from different stages of life of these men collected from nationwide social and health registers. & & & \\
 & This is a very reliable dataset with minimal measurement error and strong validity. & & & \\

\hline
YahooFinancials\footnote{\url{https://pypi.org/project/yahoofinancials/}} & This data module consists of financial data of daily stock prices of companies from the stock market. & Tim-series & Causal & Real-world 
 \\
 & Using this dataset module we can get financial data of different transactions from Yahoo Finance like stock, & & Inference & \\
 & mutual fund, cryptocurrency, ETF, US Treasury and forex for various time intervals and time ranges.& & \\

\hline

Lalonde & This is a popular observational dataset collected from the National Supported Work Demonstration. The study   & Time-series & Causal & Real-world \\
 \citep{lalonde-1} & examined how well a work training program (the treatment) affected a participant's actual wages a few years & & Inference & \\
 \cite{lalonde-2}&  after the program's conclusion. Besides the treatment indicator the dataset provides demographic variables   & & & \\
  & like age, race, academic background and previous real earnings for 260 controlled and 185 treated subjects   & & \\
  & (a total of 445) with the response (real earnings in the year 1978). In the values of the treatment assignment   & & \\
  & indicator variable 1 means treated and 0 means control/untreated. & & \\
\bottomrule
\end{tabular}
\end{table}
\end{landscape}

Besides the datasets discussed here, CauseMe \citep{CauseMe} is a public platform containing numerous ground truth causal graphs and corresponding datasets to validate the causal discovery methods. This platform contains synthetic datasets with a wide range of properties like extreme events, complex dynamics, different errors, time delay, etc. resembling real dynamical systems. Real-world datasets with multiple dimensions are also included in this benchmarking platform where the causal graph is accepted by domain experts. It is also possible to submit new time series datasets from different domains to this platform.

\subsection{Toolboxes for Causal Analysis}
Many software tools have been developed in recent years for analyzing causal relationships and effects from the dataset. We summarize some open-source causality tools with the list of available methods in Table~\ref{tab:tools}.  

\begin{landscape}
\begin{table}
\caption{Tools for performing Causal Discovery (CD) and Causal Inference (CI). Methods with (*) are suitable for time-series data.}

\label{tab:tools}
\begin{tabular}{llll}
\toprule
\textbf{Tools} &
  \textbf{Methods} &
  \textbf{Language} &
  \textbf{Category: CD / CI} \\
%
\midrule
pcalg \citep{pcalg-1} & PC, stable PC, CPC, GES, GIES, ARGES, GDS, AGES,  LINGAM, FCI, FCI–JCI, & R &
  CD, CI \\
 \citep{pcalg-2} &  FCI+,  RFCI, Generalized Adjustment Criterion (GAC),  & & \\
 & IDA algorithm, Generalized Backdoor Criterion (GBC) & & \\ 
Tetrad \citep{ramsey2018tetrad} &
  FCI, RFCI-BSC, FGES, GFCI, PC, PCStable, CPC, PcMax, RFCI, MBFS, GLASSO, FOFC, FTFC, &
  Java &
  CD \\
  &  LINGAM, TsFCI*, TsGFCI*, TsIMaGES*, MultiFASK*. & & \\ 
Causal-cmd \citep{causal-cmd} &
  FCI, FGES, Fask-Concatenated, FGES-MB, BPC, EB, Fang-Concatenated, FTFC, FAS, FOFC,  &
  Java &
  CD \\
  & GFCI, MGM, PC-ALL, PC-STABLE-MAX,  GLASSO, MBFS, IMGS\_CONT, IMGS\_DISC, &
  & \\
  & RFCI, R-SKEW, R-SKEW-E, SKEW, SKEW-E, R1, R2, R3, R4, TS-FCI*, TS-GFCI*, TS-IMGS* &
  & \\
causal-learn \citep{causallearn} &
  PC, FCI, CD-NOD, GES, ICA-based LiNGAM, DirectLiNGAM, VAR-LiNGAM* &
  Python &
  CD \\
  & RCD, CAM-UV, Additive noise models, Generalized Independence Noise (GIN) & & \\
  &  condition-based method, GRaSP, Granger causality & & \\
CausalNex \citep{CausalNex2021} &
  Bayesian Networks, Counterfactual analysis, Identify the right intervention &
  Python &
  CD, CI \\ 
LiNGAM \citep{LiNGAM} &
  DirectLiNGAM, MultiGroupDirectLiNGAM, VARLiNGAM*, VARMALiNGAM*, &
  Python &
  CD \\
  & Bootstrap, RCD, RESIT, LiM, CAM-UV, MultiGroupRCD & & \\
Tigramite \footnote{\url{https://github.com/jakobrunge/tigramite}}  &
  PCMCI*, PCMCI+*, LPCMCI*, RPCMCI*, J-PCMCI+* &
  Python &
  CD \\ 
Causal ML \citep{causalml} &
 Interaction Tree, Causal Inference Tree, Uplift Random Forests, S-Learner, T-Learner, &
  Python & CI \\
  & R-Learner, X-Learner, Doubly Robust (DR) learner, 2-Stage Least Squares (2SLS), & & 
  \\
  & Doubly Robust Instrumental Variable (DRIV) learner, DragonNet, CEVAE & & \\
EconML \citep{econml} &
  Causal Forests, Orthogonal Random Forests, S-Learner, T-Learner, X-Learner,  & Python & CI \\
  &Double Machine Learning (RLearner), Dynamic Double Machine Learning, Domain Adaptation Learner, & & \\ 
  & Doubly Robust Learner, Non-Parametric Double Machine Learning, Deep Instrumental Variables, & & \\
  & Dynamic Double Machine Learning & & \\ 
DoWhy \citep{dowhy} & Average causal effect for backdoor, frontdoor,  &
  Python &
  CI \\ 
  \citep{dowhy_gcm} & instrumental variable, and other identification methods & & \\ 
linearmodels \citep{linearmodels} &
  Fixed effects, Between estimator for panel data, First difference regression, Pooled regression for &
  Python &
  CI \\
  & panel data, Fama-MacBeth estimation of panel models, Absorbing Least Squares, Two-stage Least & & \\ 
  &  Squares, k-class Estimators, Limited Information Maximum Likelihood,  Generalized Method of & & \\ 
  & Moments, GMM estimation, Time-series estimation, 2- and 3-step estimation, Seemingly Unrelated & & \\ 
  & Regression (SUR/SURE), Three-Stage Least Squares (3SLS), Generalized Method of Moments (GMM) & & \\ 
Causal Discovery Toolbox &
  CGNN, PC, GES, GIES, LiNGAM, CAM, GS, IAMB, MMPC, SAM, CCDr &
  Python &
  CD, CI \\ 
  \citep{cdt} & ANM, IGCI, RCC, NCC, GNN, Bivariate fit, Jarfo, CDS, RECI & & \\ \bottomrule
\end{tabular}
\end{table}
\end{landscape}

\section{Conclusion}
In this survey paper, we have explored the landscape of causal discovery and causal inference methods for time-series and spatiotemporal data. We summarized common approaches, discussed limitations and assumptions, and showcased various applications across different domains of Earth science. Owing to the critical role of causal inference methods in advancing Earth science research, the review paper highlights open challenges and opportunities for causality to unravel environmental processes and ecosystem dynamics. By leveraging the methods explored in this paper, researchers can better understand the underlying mechanisms shaping Earth's systems and make informed decisions to address pressing environmental challenges. Despite their potential, we acknowledge several challenges and limitations associated with causal discovery and inference methods, including issues related to confounding, endogeneity, data stationarity and non-linearity. Addressing these challenges will require interdisciplinary collaboration, methodological innovations, and improved data integration techniques. Looking ahead, several promising avenues for future research in causal inference for Earth science emerge. Such as extending the application of causal inference methods to emerging areas of Earth system modeling, enabling methods for more robust decision-making under uncertainty, improving model interpretability and development of novel algorithms tailored to the unique characteristics of Earth science data, such as spatiotemporal dependencies, spatial interference and non-linear interactions.
By pursuing these research directions, we can further enhance the capabilities of causal discovery and inference methods to address the pressing environmental challenges facing our planet and pave the way for more sustainable and resilient Earth systems.

\section*{Acknowledgement}
This work is supported by NSF grants: CAREER: Big Data Climate Causality (OAC-1942714) and HDR Institute: HARP - Harnessing Data and Model Revolution in the Polar Regions (OAC-2118285). 

\section*{Availability Statement} No datasets were generated or analyzed during the current study. Software (other than for typesetting) was not used for this research.

\bibliographystyle{ametsocV6}
\bibliography{references}

\end{document}